\documentclass[sigconf]{acmart}

\AtBeginDocument{%
  }

\acmConference[Internetware 2025]{16th International Conference on Internetware}{June 2025}{Trondheim, Norway}

\copyrightyear{2025}
\acmYear{2025}
\setcopyright{rightsretained}
\acmConference[Internetware 2025]{the 16th International Conference on Internetware}{June 20--22, 2025}{Trondheim, Norway}
\acmBooktitle{the 16th International Conference on Internetware (Internetware 2025), June 20--22, 2025, Trondheim, Norway}
\acmPrice{}
\acmDOI{10.1145/3755930.3755977}
\acmISBN{979-8-4007-1926-4/25/06}

\usepackage{xcolor}
\usepackage{graphicx}
\usepackage{textcomp}
\usepackage{subfigure}
\usepackage{hyperref}
\usepackage{hyperxmp}
\usepackage{array,booktabs}
\usepackage{multirow}
\usepackage{balance}
\usepackage{tabularx}
\usepackage{enumitem}
\usepackage{seqsplit}

\begin{document}


\title{Brevity is the Soul of Wit: Condensing Code Changes to Improve Commit Message Generation}


\author{Hongyu Kuang} 
\affiliation{%
  \institution{State Key Lab of Novel Software}
  \institution{ Technology, Nanjing University}
  \city{Nanjing} 
  \country{China}}
\email{khy@nju.edu.cn}

\author{Ning Zhang} 
\affiliation{%
  \institution{State Key Lab of Novel Software}
  \institution{ Technology, Nanjing University}
  \city{Nanjing} 
  \country{China}}
\email{522023320204@smail.nju.edu.cn}

\author{Hui Gao} 
\affiliation{%
  \institution{State Key Lab of Novel Software}
  \institution{ Technology, Nanjing University}
  \city{Nanjing} 
  \country{China}}
\email{ghalexcs@gmail.com}
\authornote{Corresponding author}

\author{Xin Zhou} 
\affiliation{%
  \institution{State Key Lab of Novel Software}
  \institution{ Technology, Nanjing University}
  \city{Nanjing} 
  \country{China}}
\email{zhouxin@nju.edu.cn}

\author{Wesley K. G. Assunção} 
\affiliation{%
  \institution{CSC, North Carolina State University}
  \city{Raleigh} 
  \country{USA}}
\email{wguezas@ncsu.edu}

\author{Xiaoxing Ma} 
\affiliation{%
  \institution{State Key Lab of Novel Software}
  \institution{ Technology, Nanjing University}
  \city{Nanjing} 
  \country{China}}
\email{xxm@nju.edu.cn}

\author{Dong Shao} 
\affiliation{%
  \institution{State Key Lab of Novel Software}
  \institution{ Technology, Nanjing University}
  \city{Nanjing} 
  \country{China}}
\email{dongshao@nju.edu.cn}

\author{Guoping Rong} 
\affiliation{%
  \institution{State Key Lab of Novel Software}
  \institution{ Technology, Nanjing University}
  \city{Nanjing} 
  \country{China}}
\email{ronggp@nju.edu.cn}

\author{He Zhang} 
\affiliation{%
  \institution{State Key Lab of Novel Software}
  \institution{ Technology, Nanjing University}
  \city{Nanjing} 
  \country{China}}
\email{hezhang@nju.edu.cn}

\begin{abstract}
Commit messages are valuable resources for describing why code changes are committed to repositories in version control systems (e.g., Git). 
They effectively help developers understand code changes and better perform software maintenance tasks. 
Unfortunately, developers often neglect to write high-quality commit messages in practice. 
Therefore, a growing body of work is proposed to generate commit messages automatically. 
These works all demonstrated that how to organize and represent code changes is vital in generating good commit messages, including the use of fine-grained graphs or embeddings to better represent code changes. 
In this study, we choose an alternative way to condense code changes before generation, i.e., proposing brief yet concise text templates consisting of the following three parts: (1) summarized code changes, (2) elicited comments, and (3) emphasized code identifiers. 
Specifically, we first condense code changes by using our proposed templates with the help of a heuristic-based tool named ChangeScribe, and then fine-tune CodeLlama-7B on the pairs of our proposed templates and corresponding commit messages. 
Our proposed templates better utilize pre-trained language models, while being naturally brief and readable to complement generated commit messages for developers. 
Our evaluation based on a widely used dataset showed that our approach can outperform six baselines in terms of BLEU-Norm, METEOR, and ROUGE-L, with average improvements of 51.7\%, 78.7\%, and 62.5\%, respectively. 
The ablation study and human evaluation also provide further insights into the effectiveness of our approach.
\end{abstract}



 
\maketitle


\section{Introduction}

\begin{figure}[!tp]
    \centering
    \includegraphics[width=0.5\textwidth]{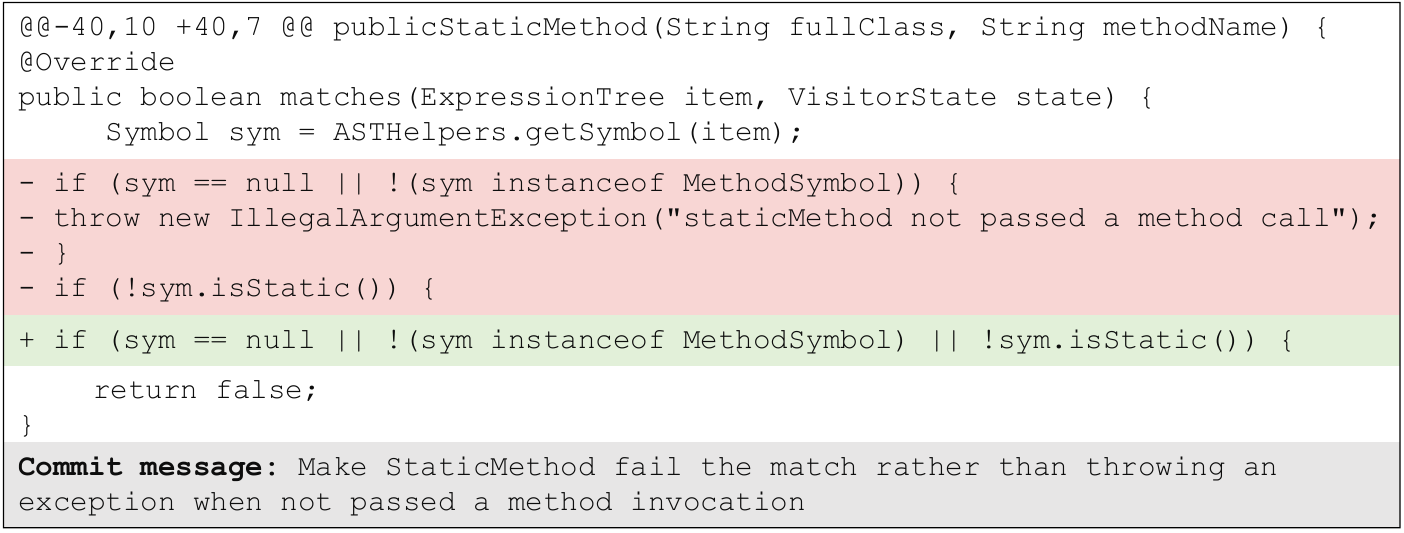}
    \caption{Code diff example}
    \label{fig:code_change_case}
\end{figure}

Online collaboration platforms, such as GitHub~\cite{GitHub} have become the vital infrastructure for modern software development.
GitHub allows intense collaborative activities with multiple developers to achieve better efficiency and fewer errors~\cite{Whitehead2007CollaborationIS}.
When developers send code changes in the repository of their project, they are suggested to also upload a brief commit message to describe the purpose of the change.\footnote{See documentation: \url{https://git-scm.com/docs/git-commit}}
These messages help developers understand the rationale behind code modifications, significantly reducing code comprehension time~\cite{DBLP:journals/ese/Kajko-Mattsson05} and facilitating software maintenance tasks~\cite{Buse_ase2010}. 
For example, Figure~\ref{fig:code_change_case} depicts a complete code change of updating a constructor method of class \texttt{RequestParams}, including related comments and annotations.
The developer also provides a concise commit message succinctly explaining that the update aims to provide the convenience of variable arguments for the constructor. 
Without the commit message depicted in the figure, other developers would have to go through each modified code statement, along with its associated comments and annotations, to infer the purpose of the changes.

Despite the importance of commit messages, developers often neglect to provide high-quality descriptions due to time constraints~\cite{DBLP:conf/icse/0001NRN13} and the overwhelming workload beyond coding tasks~\cite{DBLP:conf/msr/MaalejH10}. 
This issue is particularly prevalent in large-scale software projects ~\cite{DBLP:conf/icse/0001NRN13}.
To address the dilemma between the need for informative commit messages and developers' burdens to manually write those, a growing body of work has been proposed to automatically generate commit messages.
Early approaches relied on rule-based methods that summarize code changes through program analysis~\cite{Buse_ase2010, Cortes-Coy_2014}. 
These methods generate human-readable reports detailing what changed and where, but often failing to capture the developer’s intent and being verbose when dealing with large code modifications. 
To address this issue, Jiang et al.~\cite{DBLP:conf/kbse/JiangAM17} introduced neural machine translation (NMT) models to generate purpose-oriented commit messages by leveraging existing high-quality code commits that include both code changes and human-written messages. 
This work transformed commit message generation (CMG) into a \texttt{sequence-to-sequence (Seq2Seq)} task, leading to further proposed retrieval-based\cite{Liu_ase2018}, enhanced translation-based~\cite{Dong_icse2022}, and hybrid approaches~\cite{He_ISSTA2023}.

According to these work, \texttt{Seq2Seq} models are effective in bridging the semantic gap between code changes and their corresponding human-written descriptions.
Consequently, the representation of code changes becomes vital to their performance of generating commit messages. 
Particularly, Liu et al.~\cite{Liu_ase2018} emphasized the importance of carefully pre-processing code changes to improve generation quality. 
Follow-on approaches choose to employ fine-grained representations, such as code-change graphs~\cite{Dong_icse2022} and modification embeddings~\cite{He_ISSTA2023}, to better capture and describe code modifications. 
Recently, large language models (LLMs) are also explored to directly generate commit messages from code changes \cite{Tao_tosem2024}, while Li et al.~\cite{Li_fse2024} introduced additional change contexts as enhancements.

Despite the achieved progress, automated commit message generation still faces challenges for better performance. 
The widely used fine-grained representations, i.e., packing code changes as sequential text inputs for word embeddings~\cite{Xu_ijcai2019,Liu_msr2019,CoreGen} or composing those with augmented abstract syntax trees (ASTs) to highlight edit operations~\cite{Dong_icse2022, DBLP:conf/issta/HeWWZZ023}, often contain substantial redundant information, resulting in excessively long input sequences. 
This situation significantly increases the complexity of the task for \texttt{Seq2Seq} models because they have to first ``comprehend'' the content of changes before converting them into the rationale for modifications, trying to cover the semantic gaps between code changes and commit messages.
Moreover, the vital parts of code changes for comprehending the intents of code, such as comments and identifiers of methods and classes \cite{DBLP:journals/ese/AliSGA15}, are unfortunately not emphasized in the discussed fine-grained representations, thus also hindering the generation good commit messages.

To cope with the discussed issues, we propose our approach for commit message generation called \textbf{CONTENT} (\textbf{C}ondensing c\textbf{O}de cha\textbf{N}ges for commi\textbf{T} m\textbf{E}ssage ge\textbf{N}era\textbf{T}ion ). 
Our approach introduces a customized template as the model input, comprising three key components: (1) summarized code changes, (2) elicited comments, and (3) emphasized code identifiers. 
By transforming code changes into more comprehensible descriptions and supplementing them with explanatory comments and highlighted identifiers, our approach bridges the semantic gap between code modifications and commit messages by reducing the complexity of understanding code changes for \texttt{Seq2Seq} models, thereby enhancing the overall performance of commit message generation.


Specifically, we adapt and enhance the output of ChangeScribe \cite{ChangeScribe} to provide a concise summary of what changed and where the changes occurred, instead of using the original code \texttt{Diffs} (i.e., typically long sequences that begin with  ``-'' or ``+''). 
The summarized code changes in the template (introduced in Section \ref{sec:summarized_change}) focus on changes in classes and methods, and highlights repository names, package names, class and method names (including parameters and return value).
We then introduce change-related comments into the template as the likely intents of the developer, thus mitigating the semantic gap between implementation-oriented code changes and goal-oriented commit messages.
Finally, to further capture the implicit intents in developer-named identifiers, we use ``Camel Case'' as a conjunction to introduce emphasized code identifiers into our template with the following sequence of identifier names (if they existed in code changes): method (with extra split names), class, field, type, and others. Our comprehensive evaluation compared CONTENT against six state-of-the-art approaches of commit message generation on a widely studied benchmark\cite{liu2018neural,Xu_ijcai2019,Dong_icse2022,DBLP:journals/ijon/NieGZLLX21, DBLP:journals/tosem/WangXLHWG21,DBLP:conf/issta/HeWWZZ023}. 
Experimental results demonstrated that CONTENT achieves superior performance across all metrics, including BLEU-Norm, ROUGE-L, and METEOR. 
The ablation studies also showed that: (1) all three template components are complementary to enhance CONTENT’s performance; (2) our proposed condensed template of code changes can collaborate with both small and large language models, i.e., CodeBERT and CodeLlama, respectively. 
Furthermore, we conducted a developer-centric, human evaluation to assess the users' feedback on generated messages from CONTENT and the baselines.
We observed consistent results showing that users gave the highest scores on CONTENT’s generated commit messages, indicating that the concise yet informative summarizations of code changes (\textit{``Brevity''}) plays the key role (\textit{``the Soul''}) in generating good commit messages (\textit{``Wit''}).

In summary, this paper makes the following contributions:

\begin{itemize}
    \item \textbf{A condensed code change template} with three key components: 
    (1)~summarized code changes, 
    (2)~elicited comments, and 
    (3)~emphasized code identifiers.
    
    \item \textbf{A comprehensive experimental evaluation} comparing our approach against six state-of-the-art techniques on a widely adopted benchmark, demonstrating its effectiveness through quantitative metrics, qualitative case analyses. 
    
    \item \textbf{A developer-centric human evaluation} assessing the quality of generated commit messages, empirically validating the practical utility of our approach.
    
    \item \textbf{A publicly available package} (\url{https://github.com/huiAlex/CONTENT}) for replication and future research.
\end{itemize}


\section{Motivation}

Despite significant advancements in commit message generation, existing approaches still present several opportunities for improvement.
First, current approaches typically represent code changes either as plain text sequences or as abstract syntax trees (ASTs). However, these methods do not effectively highlight the specific differences between the original and revised code, often retaining redundant information.
Second, they often neglect valuable contextual information within code comments, which can provide essential insights into the rationale behind modifications. 
Third, identifiers such as class names and function names play a key role in shaping commit messages, but existing techniques do not explicitly account for them, leading to suboptimal results.  
To further illustrate the opportunities lying in these limitations, we present real-world examples and analysis in the following subsections.

\begin{figure}[!tp]
    \centering
    \includegraphics[width=0.48\textwidth]{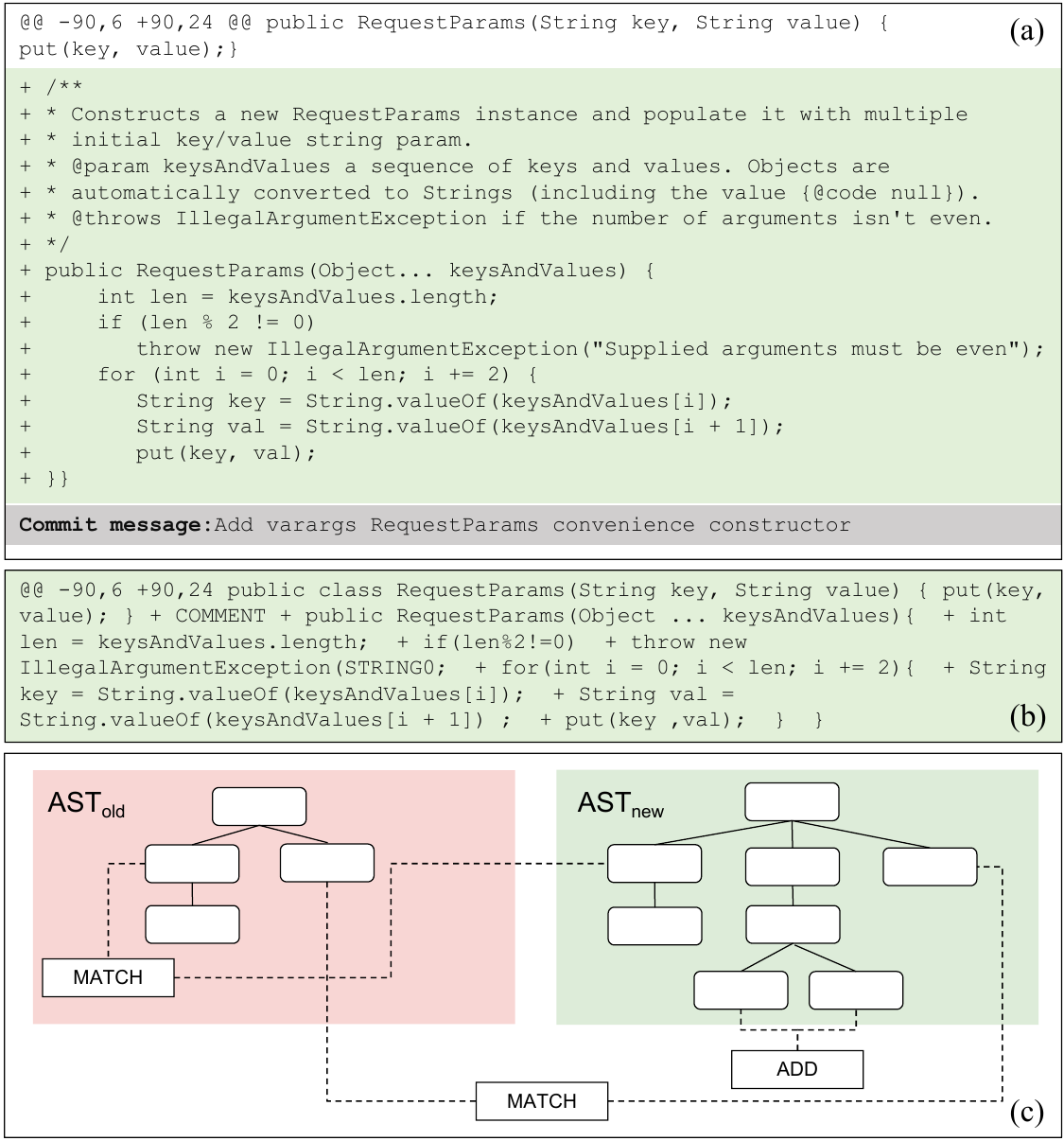}
    \caption{Different representations of code changes in existing CMG approaches: (a) git diff file; (b) original sequential order of code changes; (c) fine-grained abstract syntax trees.}
    \label{fig:motivation case}
\end{figure}

\subsection{Opportunity 1: Summarized Code Changes}
Distributed version control systems, such as Git, record code changes in diff files where modified lines are prefixed with ``-'' (indicating deletion), ``+'' (indicating addition), or remain unchanged, as illustrated in Figure \ref{fig:motivation case}(a). Existing methods process these changes using either coarse-grained or fine-grained approaches. Coarse-grained methods tokenize Git diff files while preserving the original sequence~\cite{DBLP:conf/kbse/JiangAM17,CoreGen,Liu_msr2019,Loyola_acl2017,DBLP:journals/tosem/WangXLHWG21}, occasionally inserting special tokens to differentiate between added and deleted lines~\cite{shi_emnlp2022,Xu_ijcai2019}. However, these approaches do not adequately highlight the specific differences between the old and new code, often retaining redundant information. For instance, as depicted in Figure~\ref{fig:motivation case}(b), implementation details are preserved in the representation, whereas the corresponding commit message concisely conveys the modification without extraneous details. This redundancy distracts learning-based models and degrades performance. Fine-grained approaches represent code changes using abstract syntax trees (ASTs) to emphasize edit operations through predefined special edges~\cite{Dong_icse2022}, as shown in Figure~\ref{fig:motivation case}(c). These methods employ complex architectures, such as graph neural networks, to extract features. Although they yield modest performance improvements, they introduce substantial computational overhead. Furthermore, when users make only minor updates, the input sequence must include all the information from both the old and new code hunks, thereby introducing additional redundant data. To address these challenges, CONTENT introduces a summarized representation of code changes that distills essential modification patterns, reducing model complexity and enhancing the effectiveness of commit message generation.


\subsection{Opportunity 2: Elicited Code Comments}
Commit messages are often shaped by contextual information, particularly comments that document developer intent, licensing details, and method explanations. However, recent learning-based methods, such as FIRA \cite{Dong_icse2022} and the hybrid method COME \cite{He_ISSTA2023}, overlook comment details when processing code changes. This is because both methods use javalang \cite{javalang} to generate abstract syntax trees (ASTs), which automatically replace comments with placeholders. A comparison of the outputs from NNGen \cite{Liu_ase2018} and FIRA revealed that seven of the top 15 best-performing cases in NNGen involved comment information, indicating that comments play a significant role in commit message generation. To address this, CONTENT integrates a comment module within its customized, condensed code change template. By preserving and utilizing comment information, CONTENT reduces information loss and improves the generation of more informative commit messages.

\subsection{Opportunity 3: Emphasized Code Identifiers}
Special identifiers, such as class and function names, play a central role in delineating the scope and intent of code modifications. In the CoDiSum\cite{Jiang_ase2017} dataset, which comprises 82,301 samples, 39,330 unique words were identified. Notably, only 776 words occurred more than 100 times, whereas the vast majority appeared only once or twice. These infrequent words predominantly represent user-defined identifiers, including class names, function names, and variable names.

To further examine the importance of these identifiers, we performed a syntax tree–based analysis to extract method names, class names, and variable names associated with code changes. 
We then counted their occurrence in the original commit messages and analyzed their decomposed forms to assess their prevalence. 
As shown in Table \ref{tab:identifier_frequency}, these identifiers frequently appear in commit messages, underscoring key aspects of code modifications. Their specificity not only delineates the scope of changes but also provides valuable domain-specific insights that facilitate the interpretation of modifications.

However, existing commit message generation methods do not explicitly account for these identifiers. Instead, they embed them within complex code representations, requiring models to parse intricate structures before leveraging identifier information. This indirect approach limits the ability of models to accurately capture the precise scope and intent of code changes, potentially leading to suboptimal commit message generation. To address this, we introduce a specialized identifier module within our customized condensed code changes template. This module explicitly extracts and emphasizes key identifiers, thereby enabling the model to more effectively recognize the scope and intent of code modifications. By incorporating identifier information directly into the commit message generation process, our approach enhances the accuracy and relevance of the generated commit messages.

\begin{table}[!tp]
    \centering
    \tiny

    \caption{Occurrence of user-defined identifiers in commit messages before and after splitting}
    \renewcommand{\arraystretch}{1} 
    \resizebox{\linewidth}{!}{
    \begin{tabular}{l r r}
        \toprule
        \textbf{Identifier Type} & \textbf{Occurrence Count}& \textbf{Occurrence Count After Splitting}\\ \midrule
        Method Name & 32262 & 274421 \\ 
        Class Name & 12633 & 35171 \\ 
        Variable Name & 7979 & 83144 \\ \bottomrule
    \end{tabular}}
    \label{tab:identifier_frequency}
\end{table}

\section{The CONTENT Approach}
In this section, we first introduce the condensed code changes template for CONTENT.
Then, we present the fine-tuning process applied to CONTENT.

\begin{figure}[ht]
    \centering
    \includegraphics[width=0.48\textwidth]{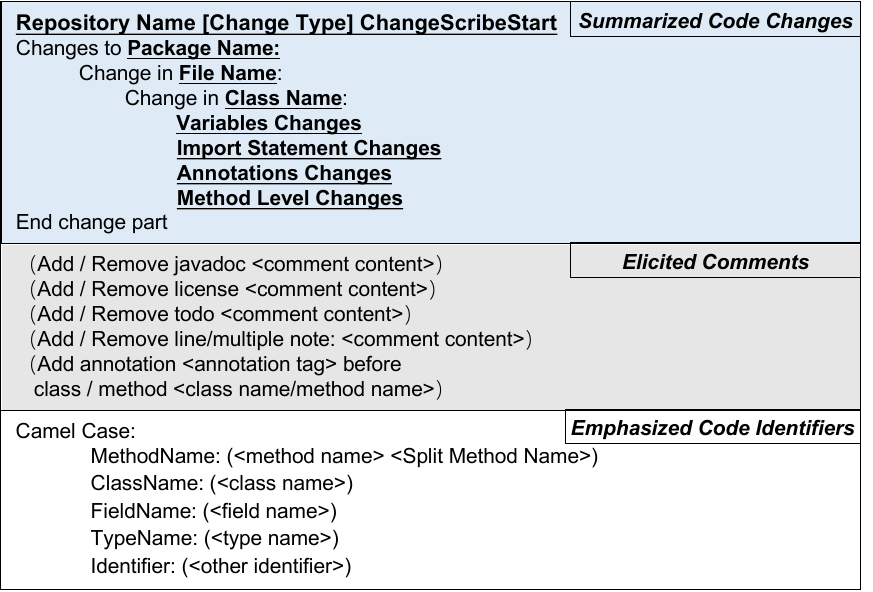}
    \caption{Code changes condensation template}
    \label{fig:template}
\end{figure}

\begin{figure}[ht]
    \centering
    \includegraphics[width=0.48\textwidth]{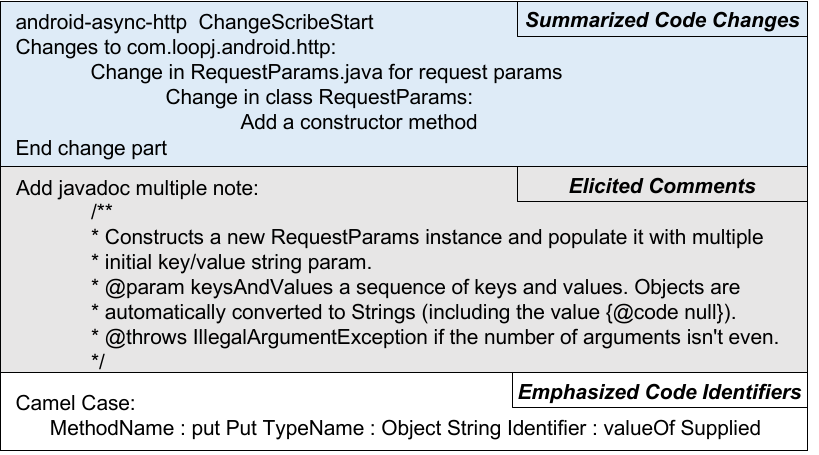}
    \caption{The template of the motivation case.}
    \label{fig: motivation case template}
\end{figure}

\subsection{Condensed Code Changes Template}
The condensed code change template is designed to capture the essential code modifications while preserving critical information for commit message generation.
Specifically, it consists of three parts as shown in Figure \ref{fig:template}: (1) Summarized Code Changes (Section \ref{sec:summarized_change}), (2) Elicited Comments (Section \ref{sec:comments}), and (3) Emphasized Code Identifiers (Section \ref{sec:identifiers}). 
Each component is described in detail, accompanied by an illustrative example of the motivation case template, as shown in Figure~\ref{fig: motivation case template}.

\subsubsection{Summarized Code Changes}
\label{sec:summarized_change}
Rather than processing raw diffs or abstract syntax trees (ASTs), CONTENT extracts key modification patterns through hierarchical summarization. This approach reduces redundancy while preserving syntactic validity, effectively addressing the information overload present in prior text-based and AST-based methods.

\begin{figure}[!tp]
    \centering
    \includegraphics[width=0.48\textwidth]{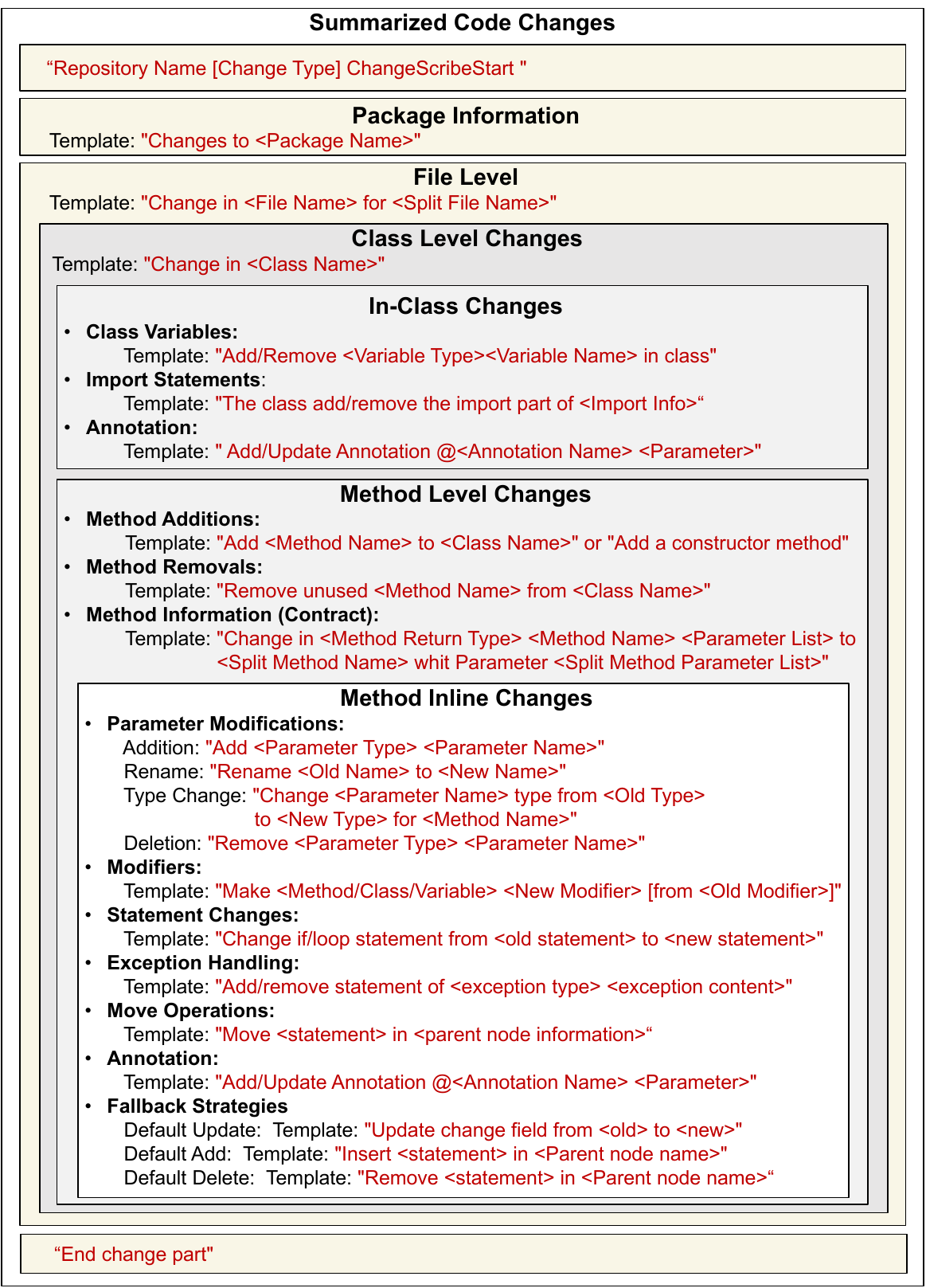}
    \caption{Detailed template of summarized code changes.}
    \label{fig: template of summarized code changes}
\end{figure}

 Figure \ref{fig: template of summarized code changes} represents the template for the Summarized Code Changes module. Instead of using the default code summary output from ChangeScribe, we introduced custom modifications to tailor it to our requirements. Following multiple experimental comparisons, we arrived at a summary template that is both readable and sufficiently concise. 

The template starts with the ChangeScribe header, generated by ChangeScribe, which includes the repository name, change type, and the “ChangeScribeStart” identifier. Table \ref{table:Change Types and Descriptions} presents 12 change types generated by ChangeScribe. If ChangeScribe fails to recognize the change type, the corresponding field is left blank, indicating an “Unknown Modification.” The header provides the model with approximate classification information, allowing it to generate commit messages that follow category-specific conventions based on the repository name and any specified change type. The template ends with the phrase “End change part.” The details of each component of the template are explained in the following sections.

\begin{table}[ht]
\centering
\caption{Change types in ChangeScribe}
\resizebox{\linewidth}{!}{
\begin{tabular}{l l l }
\toprule
\textbf{Change Type}&\textbf{Description}&\textbf{Identifier}\\
    \midrule    
    Structure Modification & Changes involving setter/getter methods for simple data access/modification. & Ty0 \\
    \midrule
    State Access Modification & Accessor methods providing data without modifying members. & Ty1 \\
    \midrule
    Update Modification & Mutator methods modifying object state. & Ty2 \\
    \midrule
    Behavior Modification & Methods implementing complex internal behavioral changes. & Ty3 \\
    \midrule
    Object Creation Modification & Object creation/destruction methods (constructors, factories). & Ty4 \\
    \midrule
    Relationship Modification & Methods establishing class relationships (generalization, association). & Ty5 \\
    \midrule
    Control Modification & Methods modifying external class behavior (controllers, factories). & Ty6 \\
    \midrule
    Large Modification & Commits combining multiple roles with numerous methods. & Ty7 \\
    \midrule
    Lazy Modification & Predominantly getter/setter methods indicating incomplete features. & Ty8 \\
    \midrule
    Degenerate Modification & Empty/abstract methods signaling planned features. & Ty9 \\
    \midrule
    Small Modification & Minor changes without significant system impact. & Ty10 \\
    \midrule
    Unknown Modification & Unclassified changes with minimal system impact. & Ty11 \\

\bottomrule
\end{tabular}}
\label{table:Change Types and Descriptions}
\end{table}
                    
\textbf{a. Package Name}:
                Packages organize code logically, grouping classes and interfaces. The output template for package names aims to indicate the precise location of changes in the codebase, formatted as follows: \texttt{Changes to \textless Package Name\textgreater}.
                
\textbf{b. File Name}:
                A single code change can involve multiple files, each indicated by a delimiter. We selected “change” as the verb to represent modifications and used “in” to specify the location, followed by the details for each file.

\textbf{c. Class Name}: 
                A single modified file may contain multiple classes, including inner classes. The template for each class is \texttt{Change in <Class Name>}. If the file contains only one class, the class name is omitted because the file name typically matches the class name.
                
\textbf{d. In-Class Changes}: In-class changes primarily involve adding or removing class variables, modifying import statements, and updating class-level annotations. These modifications are captured by three templates, as illustrated in the “In-Class Changes” section of Figure~\ref{fig: template of summarized code changes}.
    
\textbf{e. Class Method Additions and Removals}: Method changes typically reflect business or refactoring requirements and often indicate the commit’s primary intent. As shown in the “Method Level Changes” section of Figure~\ref{fig: template of summarized code changes}, two templates are defined to capture these changes.
    
\textbf{f. Class Method Information}: A method’s return type, name, and parameters together define its “contract,” with each element specifying a distinct aspect of the method’s behavior: \begin{itemize} \item \textbf{Return Type}: Outlines the expected outcome. \item \textbf{Method Name}: Summarizes the method’s functionality. \item \textbf{Parameter List}: Describes the required inputs. \end{itemize} We capture these components, as illustrated in the “Method Level Changes” section of Figure~\ref{fig: template of summarized code changes}. 
    
\textbf{g. Method Inline Changes}: In the ``Method Inline Changes'' section, we reused several rule templates from ChangeScribe. To better align the output with conventional commit message patterns, we customized and modified these templates to emphasize our primary points of interest. The final templates are illustrated in Figure~\ref{fig: template of summarized code changes} under the ``Method Inline Changes'' section. Specifically, the modifications include:\begin{itemize}
        \item \textbf{Parameter Modifications:} Adjustments to parameters to reflect functional changes.
        \item \textbf{Modifiers:} Modifiers such as \textit{private} or \textit{public} specify access levels, while the term ``make'' is used to indicate visibility adjustments.
        \item \textbf{Statements:} Modifications to branch and loop statements are significant \cite{van2019generating}, as they capture differences between the pre- and post-modification states.
        \item \textbf{Exceptions:} Enhancements in exception handling improve robustness.
        \item \textbf{Move Operations:} ChangeScribe’s categories for additions, deletions, and modifications were expanded to include move operations based on practical testing needs.
        \item \textbf{Annotations:} Method-level annotations often contain important information and guide the model in generating more precise commit messages.
        \item \textbf{Fallback Strategies:} Three fallback strategies are employed to ensure comprehensive coverage of other changes.
    \end{itemize}


The summarized code changes for the motivation case, highlighted in blue in Figure \ref{fig: motivation case template}, eliminate redundant information irrelevant to commit message generation. This approach effectively narrows the semantic gap between the code changes and the corresponding commit messages.

\subsubsection{Elicited Comments}
\label{sec:comments}
While ChangeScribe can extract inline comments, it does not handle class- or method-level comments, which often contain critical explanations, legal notices, or TODO items. These comments frequently capture the rationale behind code modifications, making them vital for commit message generation.
Therefore, we employ regular expressions to extract these Comments Overlooked by ChangeScribe. As shown in Figure \ref{fig:template}, we propose categorizing comments into four types:

    (1) \textbf{JavaDoc Comments}: Multiline comments with more than 20 tokens. 

    (2) \textbf{License Comments}: Comments containing the keyword ``license," often used to indicate licensing terms. 
    
    (3) \textbf{TODO Comments}: Comments containing the keyword ``todo," representing pending tasks or planned modifications. 
    
    (4)  \textbf{General Comments}: Comments providing method explanations, algorithm descriptions, or additional context. 
    
Beyond comments, annotations\footnote{https://docs.oracle.com/javase/tutorial/java/annotations/index.html} play a crucial role in documenting and enhancing code functionality. For example, in frameworks such as Spring\footnote{https://docs.spring.io/spring-framework/reference/testing/annotations.html}, the \texttt{@SqlConfig} \cite{SqlConfig} annotation specifies how SQL scripts should be parsed and executed, while the annotation \texttt{@TestPropertySource} \cite{TestPropertySource} configures test properties for a test class.
These annotations provide essential context and functionality integral to code behavior. However, the current ChangeScribe tool ignores certain annotations deemed unimportant. To address this, we use regular expressions to extract these annotations and propose a dedicated template for capturing annotation changes, ensuring they are accurately documented. This enhancement facilitates a more comprehensive understanding of code modifications and improves overall maintainability.
Figure \ref{fig:template} displays the elicited comments template, whereas the corresponding elicited comments for  motivation case are highlighted in gray color in Figure \ref{fig: motivation case template}.



\subsubsection{Emphasized Code Identifiers}
\label{sec:identifiers}
  In software development, identifiers such as variables, function names, and class names convey significant semantic meaning through their CamelCase naming conventions. To enhance semantic utilization in commit message generation, we systematically categorize identifiers into five semantic types: MethodName, ClassName, FieldName, TypeName, and other identifiers (e.g., annotations, enumeration values). The ordering of these categories reflects both empirical observations and code structure conventions: method and class names are often central to commit intent, whereas field and type names typically play supporting roles. This finding aligns with studies demonstrating that detailed identifiers yield more informative and context-aware commit messages \cite{DBLP:conf/icse/DongLHT23}. By prioritizing identifiers, we emphasize the most salient semantic units for message generation. Moreover, the substantial number of identifiers necessitates filtering and refinement. 
  We eliminate identifiers that appear frequently in the code but rarely in commit messages and format the remaining identifiers to strictly conform to CamelCase conventions.
  Figure \ref{fig:template} displays the emphasized identifiers template, whereas the corresponding emphasized  identifiers for motivation case are highlighted in white color in Figure \ref{fig: motivation case template}.


\subsection{Supervised Fine-Tuning for CONTENT}

Since supervised fine-tuning (SFT) is the mainstream approach in the LLMs training \cite{DBLP:journals/corr/abs-2403-13372}, for open-source LLMs (e.g., LLaMA-2 \cite{touvron2023llama}, Gemma \cite{DBLP:journals/corr/abs-2403-08295}), the most straightforward method to enable the model to adapt a specific domain quickly is to use collected domain label to perform SFT on the model. 
The SFT phase is typically the preliminary phase of the well-designed training framework \cite{DBLP:conf/nips/RafailovSMMEF23}, as well as the fine-tuning of commit message generation. 
The auto-regressive generation process of commit message $Y$ can be formulated as follows:
\begin{equation}
    P_\pi(Y \mid \mathcal{P}) = \prod_{k=1}^n P_\pi(y_k \mid \mathcal{P}, Y_{1:k-1}),\tag{1}
    \label{eq:sft}
\end{equation}

where $Y = \{y_1, y_2, \dots, y_n\}$ is the commit message of length $n$, $y_k$ is the corresponding $k^\text{th}$ token of the commit message, $Y_{1:k-1}$ is the prefix sequence of $Y$ ahead the token $y_k$. $P_\pi(y_k \mid \cdot)$ is a conditional probability of a LLM $\pi$ for generating the $k^\text{th}$ token of $Y$ based on the input prompt $\mathcal{P}$ and the prefix sequence.

Given a basic open-source model $\pi^0$, the goal of SFT is to obtain a model $\pi^{SFT}$ through minimizing the cross-entropy loss:
\begin{equation}
\mathcal{L}_{SFT} = -\sum_{k=1}^n \log P_{\pi^0}(\hat{y}_k = y_k \mid \mathcal{P}, Y_{1:k-1}),\tag{2}
\label{loss function}
\end{equation}
where $\hat{y}_k$ is the $k_{th}$ token of the generated commit message $\hat{Y}$, and $Y$ is the corresponding ground-truth label.

In our method, we utilize the CodeLlama-7B model \cite{CodeLlama} as the base, which has been pre-trained on a large corpus of code datasets spanning various tasks. 
This pre-training enables the model to effectively capture general language structures, making it highly suitable for generating commit messages from concise code changes. 
The model is fine-tuned on a dataset \( T = \{(y_i, \mathcal{P}_i)\} \), where \( \mathcal{P}_i \) represents the formatted and prompted condensed code changes, and \( y_i \) denotes the corresponding commit message. The loss function \ref{loss function} is designed to encourage the model to produce commit messages that closely align with the ground truth.

\section{Experiment Setup}
\label{sec:setup}

To evaluate CONTENT, we conduct an empirical assessment with four primary objectives: 
(1) to determine whether the utilization of concise text templates of code changes could enhance the performance of automatic commit message generation, (2) to analyze the impact of the different parts of concise text templates on the performance of CONTENT, (3) to explore if concise text template can be applied to the different pre-trained models effectively, and (4) to further evaluate the quality of commit messages generated by CONTENT and the state-of-the-art through an extensive human evaluation. 
We formulate the following four RQs:



\vspace{3mm} \noindent
\textbf{RQ1: Overall effectiveness.} \textit{How does CONTENT perform compared to the state-of-the-art commit message generation techniques?} 

\vspace{3mm} \noindent
\textbf{RQ2: Ablation analysis.} \textit{How does each component of CONTENT template contribute to the effectiveness?}

\vspace{3mm} \noindent
\textbf{RQ3: Generalization capability.} \textit{How does condensed code changes template of CONTENT perform when combined with different code pre-trained models?} 

\vspace{3mm} \noindent
\textbf{RQ4: Human evaluation.} \textit{How does CONTENT perform from the perspective of developers?}

\subsection{Dataset}
 We used a widely recognized dataset in this field, initially collected by Jiang et al.\cite{Jiang_ase2017} from the top 1000 starred Java projects on GitHub. 
 As subsequent studies identified quality issues in some commit messages within this dataset, including irrelevant and auto-generated messages, we opted for the high-quality, filtered dataset provided by Xu et al.\cite{Xu_ijcai2019}. 
 To meet ChangeScribe's input requirements, we retrieved the original code version pairs and associated repositories. 
 Due to the dataset's age and project maintenance issues, we ultimately obtained 87,602 usable data pairs from an initial set of 90,661.

\subsection{State-of-the-art baselines}
  CONTENT is compared against six state-of-the-art techniques for commit message generation.
  
  \textbf{Information retrieval-based techniques} leverage information retrieval (IR) methods to identify and reuse existing commit messages from similar code changes. NNGen\cite{liu2018neural} serves as the representative IR-based technique in this comparison.
  
  \textbf{Learning-based techniques} employ neural machine translation (NMT) models to automatically generate commit messages. The four learning-based techniques included in the comparison are CODISUM\cite{Xu_ijcai2019}, FIRA\cite{Dong_icse2022} and Coregen\cite{DBLP:journals/ijon/NieGZLLX21}. 
  
  \textbf{Hybrid approaches } generate commit messages by integrating information retrieval methods with neural machine translation algorithms. CoRec\cite{DBLP:journals/tosem/WangXLHWG21} and , and COME\cite{DBLP:conf/issta/HeWWZZ023} are the hybrid approaches considered in this study.


\subsection{Implementation}

We adopt a learning-based approach using the pre-trained CodeLlama-7B model, designed for general code synthesis and understanding \cite{CodeLlama}. 
To further assess the generalizability of our method, we extend our experiments to include CodeBERT.

For CodeLlama-7B, input data is preprocessed to meet the model’s requirements. The model accepts two input types: code\_tokens (tokenized code change data) and docstring\_tokens (tokenized commit messages). 
The code change data undergoes punctuation handling, word splitting, and tokenization, while commit messages are similarly preprocessed and tokenized. 
Additionally, a \textless reponame, hashcode\textgreater pair is generated for each entry to link it to the original code change, enabling efficient comparison and tracking.
The maximum input length of CodeLlama-7B is set to 1024 tokens, while commit messages are limited to 128 tokens due to their concise nature. 
The model is fine-tuned using the LoRA (Low-Rank Adaptation) technique with a batch size of 2, a learning rate of $1 \times 10^{-4}$, and 3 epochs.

For CodeBERT, a similar preprocessing pipeline is applied. 
The fine-tuning parameters are as follows: learning rate = $5 \times 10^{-5}$, batch size = 32, beam size = 10, source length = 256, target length = 128, and 10 epochs.

All experiments are conducted on a server running Ubuntu 22.04 with an Intel Xeon(R) Gold 6130 CPU @ 2.10GHz. 
The models are trained on a single NVIDIA Tesla V100 GPU (32GB memory), with each training session taking approximately 12 hours.


\subsection{Evaluation Metrics}
The manually written commit messages from the dataset serve as ground truth. For each code change, the similarity between the generated and ground-truth commit messages is evaluated. Following prior research on commit message generation \cite{Liu_ase2018,Liu_tse2022,Hoang_icse2020,Wang_tosem2021,Xu_ijcai2019}, we use the established similarity metrics BLEU-Norm, ROUGE-L, and METEOR.
\begin{itemize}
    \item \textbf{BLEU-Norm} is a case-insensitive variant of BLEU that calculates precision by averaging modified n-gram precisions across 1- to 4-grams (BLEU-4) \cite{acl/PapineniRWZ02}. It measures matching n-grams as a proportion of total n-grams in the sequence. A recent study \cite{icsm/TaoWSDH0ZZ21} found that BLEU-Norm aligns most closely with human judgment in commit message quality assessment, making it a suitable evaluation metric.
    \item \textbf{ROUGE-L} determines the F-score based on the longest common subsequence (LCS) between the generated and ground-truth sequences\cite{conf/iwpc/LiNJWHW18}. Longer LCS values indicate greater similarity between sequences.
    \item \textbf{METEOR} evaluates precision and recall at the unigram level, computing their harmonic mean while applying a penalty for non-adjacent matching tokens \cite{METEOR}.
\end{itemize}

\section{RESULTS AND ANALYSIS}
In this section, we present the overall results of CONTENT (RQ1) in Section \ref{sec:rq1}, followed by the results of the ablation study (RQ2) in Section \ref{sec:rq2}, the evaluation of generalization capability (RQ3) in Section \ref{sec:rq3}, and the human evaluation (RQ4) in Section \ref{sec:rq4}.
\begin{table}[!bp]\footnotesize
\caption{Overall commit message generation results}
\renewcommand{\arraystretch}{0.9}
\tiny
\resizebox{\linewidth}{!}{
\begin{tabular}{l|rrr}
\toprule
\textbf{Approach}   & \textbf{BLEU-Norm}         & \textbf{METEOR}       & \textbf{ROUGE-L} \\ 
\midrule
NNGen     & 9.01(+136\%)& 9.06(+177\%)& 11.09(+153\%)\\
CoDiSum & 16.43(+29.3\%)& 16.58(+51.4\%)& 19.62(+42.8\%)\\
CoRec     & 12.89(+64.8\%)& 12.78(+96.5\%)& 15.34(+82.7\%)\\
Coregen     & 14.08(+50.9\%) & 13.80(+82.0\%)& 18.14(+54.5\%)\\
FIRA     & 17.57(+20.9\%)& 18.51(+35.7\%)& 21.48(+30.4\%)\\
COME     & 19.50(+8.9\%)& 21.93(+29.4\%)& 24.92(+12.4\%)\\ \midrule
CONTENT    & \textbf{21.24} & \textbf{25.11}& \textbf{28.02}\\ 
Avg.         &+51.7\%  &+78.7\%   &+62.5\% \\
\bottomrule
\end{tabular}}
\label{tab:rq1}
\end{table}

\subsection{RQ1: Overall Effectiveness}
\label{sec:rq1}
Table \ref{tab:rq1} presents the average scores for BLEU-Norm, METEOR, and ROUGE-L across all commit messages generated by the baseline approaches and CONTENT. 
CONTENT consistently outperforms all baseline approaches, including the strongest IR-based method (NNGen) and the state-of-the-art learning-based model (COME), across all metrics.  
This demonstrates CONTENT’s effectiveness in both precision (BLEU-Norm, ROUGE-L, METEOR) and recall (ROUGE-L, METEOR).
Specifically, CONTENT achieves improvements ranging from 8.9\% to 136\% (51.7\% on average) in BLEU-Norm, 29.4\% to 177\% (78.7\% on average) in METEOR, and 12.4\% to 153\% (62.5\% on average) in ROUGE. 
Notably, CONTENT shows substantial gains over NNGen, with increases of 136\% in BLEU-Norm, 177\% in METEOR, and 153\% in ROUGE-L. 
When compared to COME, CONTENT achieves an average improvement of 16.9\% across the three metrics.

\begin{figure}[!tp]
    \centering    \includegraphics[width=0.48\textwidth]{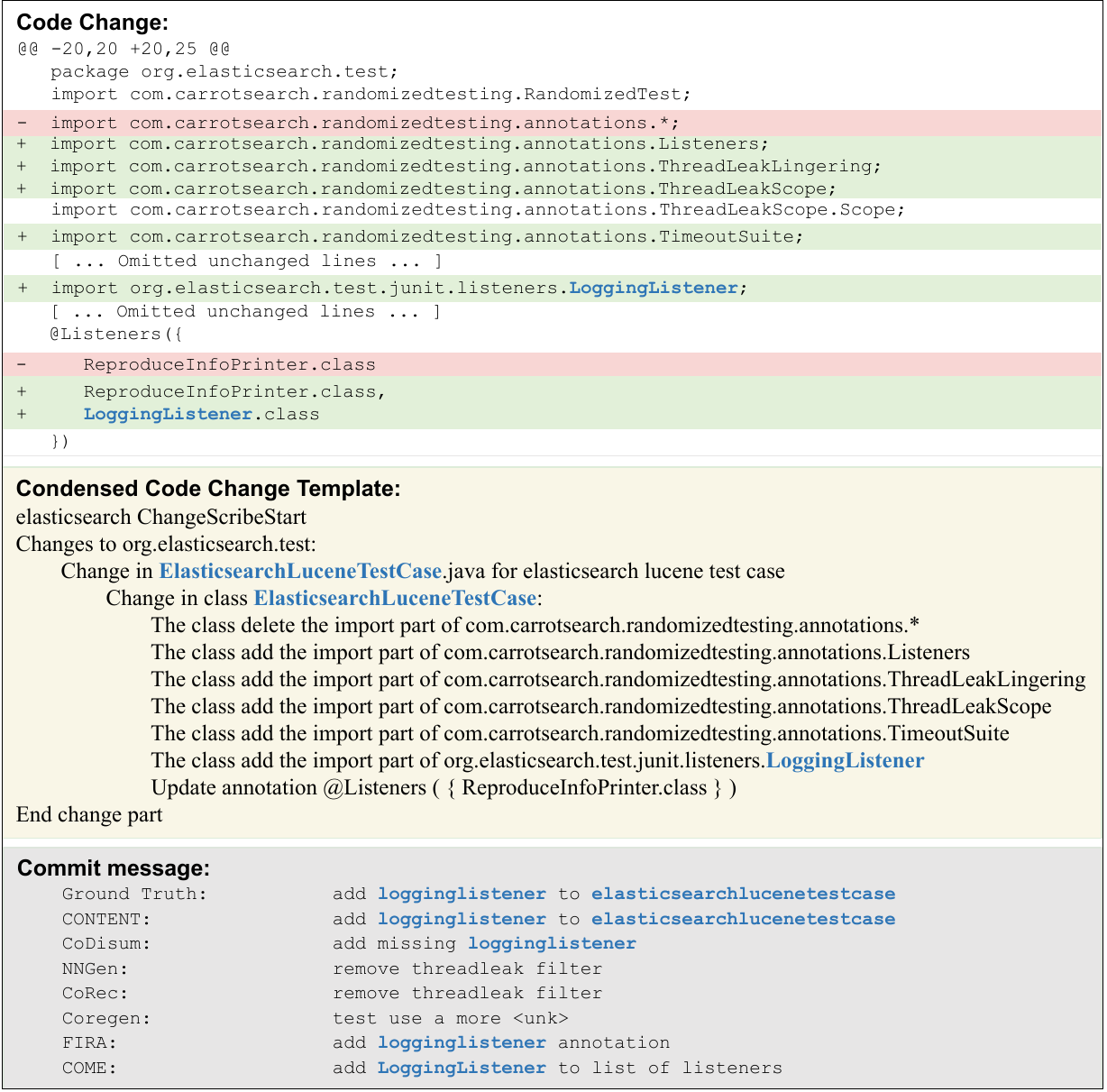}
    \caption{the effectiveness of summarized code changes}
    \label{fig:rq1}
\end{figure}

\sloppy
These improvements are attributed to CONTENT’s condensed code changes template, which capture key code modifications more effectively than other approaches.
For instance, as shown in Figure \ref{fig:rq1}, a commit modifies the file \texttt{ElasticsearchLuceneTestCase.java} by adding a \texttt{LoggingListener} class. 
CONTENT effectively summarizes these key changes, whereas other methods struggle due to their limitations.
COME, CoDisum, and FIRA fail to incorporate essential information such as file names, leading to commit messages that omit key context (e.g., ``\texttt{ElasticsearchLuceneTestCase}'').
As retrieval-based methods, NNGen and CoRec struggle to generate relevant commit messages when the dataset lacks similar prior commits, producing outputs that do not accurately reflect the actual changes.
Since CoreGen relies on contextual signals, it struggles with minor code changes, leading to less precise commit messages.
In contrast, CONTENT produces more precise and concise commit messages by effectively incorporating critical contextual elements into the template, resulting in significantly improved performance over existing methods.

\subsection{RQ2: Ablation Study on CONTENT} 
\label{sec:rq2}
To assess the effectiveness of each component in the CONTENT template, we conducted an ablation study. 
The core innovation of CONTENT lies in its approach to condensing code changes through a structured template composed of three key elements: summarized code changes, elicited comments, and emphasized code identifiers. 
To examine the impact of each component, we designed three CONTENT variants: (1) removing elicited comments from the template (denoted as CONTENT$_{comm-}$), (2) removing emphasized code identifiers from the template (denoted as CONTENT$_{ident-}$), and (3) directly using raw code changes instead of using summarized ones (denoted as CONTENT$_{summ-}$).

Table \ref{tab:rq2} shows the performance of the original CONTENT template alongside these variants, allowing us to analyze both the quantitative and qualitative contributions of each component.

\begin{table}[!tp]
\caption{The ablation study on CONTENT (\textit{comm-}: without elicited comments, \textit{ident-}: without emphasized identifiers, and \textit{summ-}: using raw diff instead of summarized changes)}
\label{tab:rq2}
\renewcommand{\arraystretch}{0.8}
\tiny
\resizebox{\linewidth}{!}{
\begin{tabular}{l|ccc}
\toprule
\textbf{Approach}   & \textbf{BLEU-Norm}         & \textbf{METEOR}       & \textbf{ROUGE-L} \\ \midrule
CONTENT$_{comm-}$ & 20.68 & 23.87 & 26.84\\
CONTENT$_{ident-}$ & 20.99 & 24.38 & 27.28\\ 
CONTENT$_{summ-}$  & 20.83  & 24.71 & 27.69\\
\midrule
CONTENT    & \textbf{21.24} & \textbf{25.11}& \textbf{28.02}\\ 
\bottomrule
\end{tabular}}
\end{table}

\subsubsection{Contribution of Elicited Comments}

\begin{figure}[ht]
    \centering
    \includegraphics[width=0.48\textwidth]{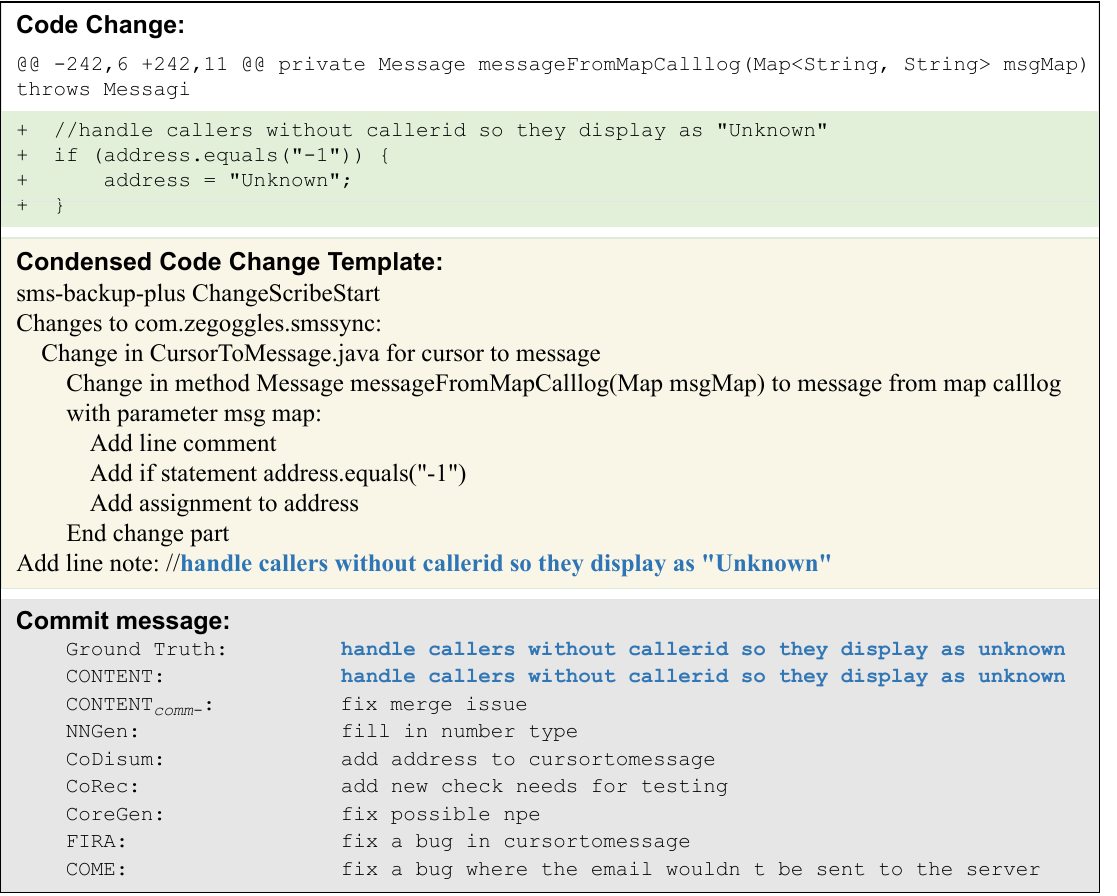}
    \caption{Case analysis: elicited comments}
    \label{fig:Case analysis: elicited Comments}
\end{figure}

As demonstrated in Table~\ref{tab:rq2}, CONTENT$_{comm-}$ consistently underperforms across all evaluated metrics, suggesting that elicited comments improve the quality of generated commit messages. 
To explore this further, we analyzed instances where CONTENT$_{comm-}$ yielded inferior results compared to the default CONTENT. 
Figure \ref{fig:Case analysis: elicited Comments} provides a representative example, detailing code modifications, the ground truth commit message, and outputs from default CONTENT, CONTENT$_{comm-}$, and other baseline models. In this example, the developer included a line comment in the code: ``handle callers without callerid so they display as unknown.'' 
The elicited comments embedded in the CONTENT template enabled the default CONTENT to generate a commit message identical to the developer’s ground truth. 
Conversely, CONTENT$_{comm-}$ produced a less precise message. 
Additionally, other evaluated methods, such as NNGen, CoRec, and CoreGen, generated entirely unrelated commit messages, while CODISUM, FIRA, and COME focused solely on code changes, overlooking comment content. 
This limitation arises because these methods either treat comments as standard code changes or retain only the action of adding a comment (e.g.,``+ SINGLE'') without preserving its substantive details, hindering the model’s ability to leverage valuable commit message information embedded in comments during training.

\subsubsection{Contribution of emphasized code identifiers.}
\begin{figure}[hb]
    \centering
    \includegraphics[width=0.48\textwidth]{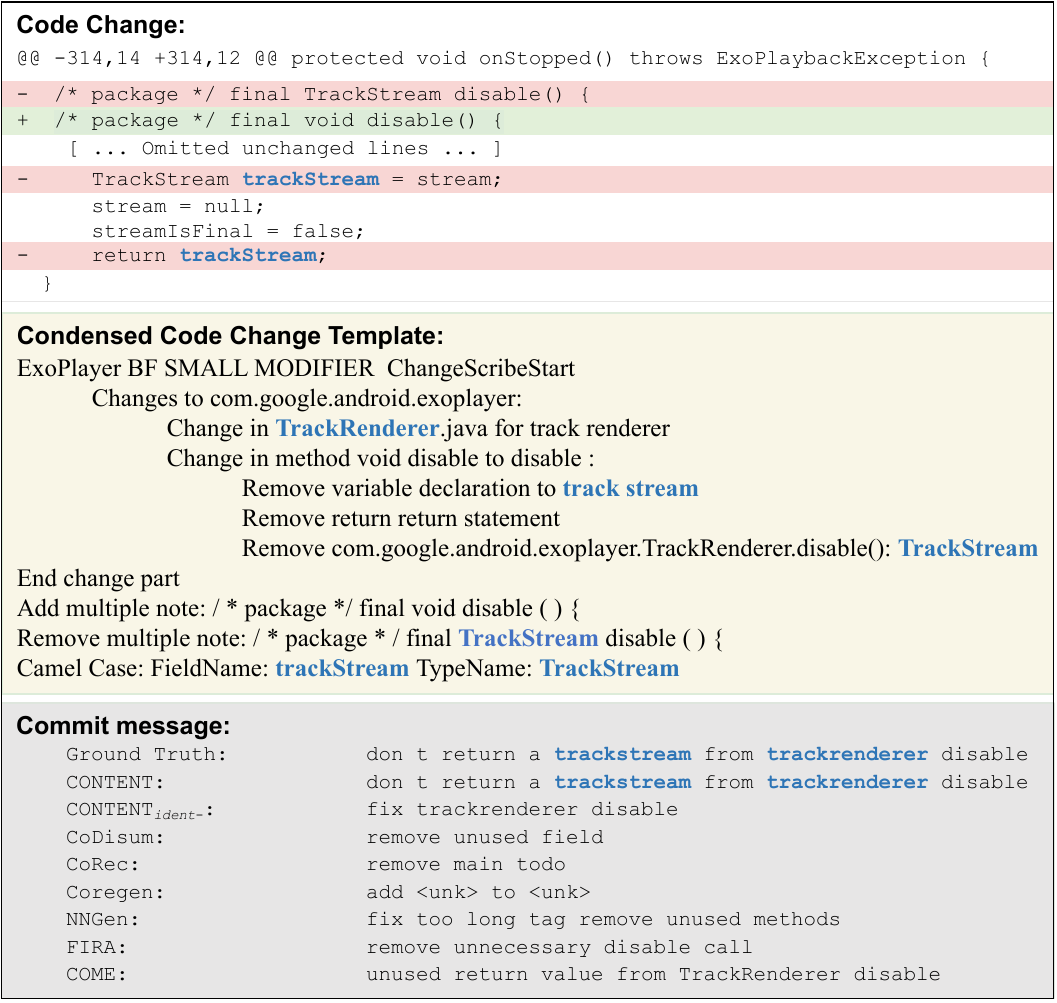}
    \caption{Case analysis: emphasized code identifiers}
    \label{fig:Case analysis: emphasized code identifiers}
\end{figure}

As shown in Table \ref{tab:rq2}, CONTENT$_{ident-}$ also exhibits a notable decline in performance, confirming that emphasized code identifiers aid in commit message generation. 
To understand this impact, we analyzed cases where CONTENT$_{ident-}$ performed worse than the default CONTENT.
Figure \ref{fig:Case analysis: emphasized code identifiers} presents a real-world example from our dataset, displaying code changes, the condensed template, ground truth, and outputs from various methods.
In this case, the developer's commit message included a field name from the modified code: ``\texttt{trackstream}''. 
The default CONTENT successfully generated an identical commit message by leveraging emphasized code identifiers. 
However, CONTENT$_{ident-}$, which lacks these identifiers, failed to include the uncommon field name in its output. 
Notably, other comparative techniques also struggled with this example: 
NNGen produced an entirely irrelevant commit message.
FIRA failed to accurately describe the change.
COME performed better but still omitted ``trackstream'', likely because it is a rare identifier not frequently appearing in the training set.

\subsubsection{Contribution of Summarized Code Changes}
As shown in Table \ref{tab:rq2}, CONTENT$_{summ-}$ consistently underperforms compared to the default CONTENT across all evaluated metrics, highlighting the significance of summarized code changes in enhancing commit message quality.
For instance, in the code change illustrated in Figure \ref{fig:rq1}, CONTENT$_{summ-}$ produces the output ``add logginglistener to lucene based tests,'' suggesting that the summarized code change module effectively captures change path information. 
In contrast, without this module, the model—constrained by a limited training dataset—struggles to accurately delineate the scope of changes. This difficulty is further exacerbated by the presence of substantial redundant information in the code diff that is unrelated to the commit message, leading to incomplete commit messages.

\vspace{+4mm}
This ablation study confirms that each component of the CONTENT template plays a significant role in enhancing the quality of generated commit messages, particularly by ensuring the relevance and precision of the commit messages to the exact code changes.

\subsection{RQ3: Generalization Capability}
\label{sec:rq3}
To evaluate whether CONTENT’s template can be effectively integrated with different code pre-trained models, we selected two state-of-the-art models: CodeBERT \cite{Feng_emnlp20} and CodeLlama-7B \cite{CodeLlama}. 
They are widely recognized for their applications in code comprehension \cite{Shang_icsme2024, Hu_ase2024, Ma_tosem2024} and code generation \cite{Qu_ASE2024, Yang_tse2024, Weyssow_tosem2025}.  
CodeBERT is an \textit{encoder-only} model based on RoBERTa \cite{liu2019roberta}, designed for code understanding and representation learning. 
CodeLlama-7B, a \textit{decoder-only} model built on Llama 2 \cite{touvron2023llama}, is pre-trained on large-scale code datasets and optimized for code-related tasks.
Notably, CodeLlama-7B is classified as a LLM with 7 billion parameters, whereas CodeBERT, with approximately 125 million parameters, is considerably smaller and tailored for tasks requiring deep code comprehension capabilities.
These models represent distinct architectures for code understanding, making them well-suited for evaluating CONTENT’s adaptability across different pre-trained approaches.

As shown in Table \ref{tab:rq3}, the experimental results demonstrate that incorporating the CONTENT template into different pre-trained models leads to significant improvements across all evaluation metrics. 
When combined with CodeBERT, the BLEU-Norm score increased from 18.13 to 19.15, the METEOR score improved from 19.97 to 21.14, and the ROUGE-L score rose from 21.72 to 23.06. 
Similarly, when integrated with CodeLlama-7B, the BLEU-Norm score increased from 20.63 to 21.24, the METEOR score improved from 24.29 to 25.11, and the ROUGE score rose from 27.44 to 28.02.
On average, the CONTENT template improves performance by 4.3\% (BLEU-Norm), 4.65\% (METEOR), and 4.15\% (ROUGE-L) across different models. 
Additionally, the results indicate that the inherent strengths of the base model significantly influence the overall performance of the method. When CodeLlama is used as the base model, its performance surpasses all baseline methods.
These results highlight the template’s effectiveness in capturing and conveying key information in code changes, enhancing the understanding and generation capabilities of various pre-trained models. 
Consequently, CONTENT demonstrates robust generalization across different learning architectures, making it a versatile tool for improving commit message generation.

\begin{table}[t]\footnotesize
\caption{Results of combined with different code pre-train models}
\label{tab:rq3}
\centering
\resizebox{\linewidth}{!}{
\begin{tabular}{l|rrr}
\toprule
\textbf{Approach}   & \textbf{BLEU-Norm}         & \textbf{METEOR}       & \textbf{ROUGE-L} \\ 
\midrule
CodeBERT    & 18.13(+5.6\%)& 19.97(+5.9\%)& 21.72(+6.2\%)\\
CONTENT$_{CodeBERT}$    & 19.15& 21.14& 23.06\\\midrule
CodeLlama-7B     & 20.63(+3.0\%)  & 24.29(+3.4\%)& 27.44(+2.1\%)\\
CONTENT$_{CodeLlama-7B}$    & \textbf{21.24} & \textbf{25.11}& \textbf{28.02}\\ 

\bottomrule
\end{tabular}}
\end{table}

\subsection{RQ4: Human Evaluation}
\label{sec:rq4}
Automated evaluation metrics measure text similarity between generated and reference texts, but they fail to account for syntactic, grammatical, and structural features. Consequently, we conducted a human evaluation to further assess CONTENT's practical utility. Specifically, we designed a study evaluating commit message quality from developers' perspectives, comparing CONTENT's output against messages generated by the top retrieval-based method (NNGen), the leading learning-based technique (FIRA), and the optimal hybrid approach (COME). Six industrial Java developers\footnote{None are co-authors of this paper.} with 2–8 years of experience participated.

\subsubsection{Study Design}
Following previous work\cite{DBLP:conf/icsm/Liu0K0BT18, Wang_tosem2021, Dong_icse2022}, we randomly selected 100 commits from the test set and designed a questionnaire for manual evaluation. 
For each commit, the questionnaire included the code change, the ground truth commit message, and the commit messages generated by CONTENT, NNGen, FIRA, and COME. 
Participants were asked to score these messages on a scale from 0 to 4, where higher scores indicated greater semantic similarity to the ground truth message. 
We followed the scoring criteria from previous studies \cite{DBLP:conf/icsm/Liu0K0BT18, Dong_icse2022}, detailed in Table \ref{tab:scoring_criterion}.
To mitigate bias, the techniques were anonymized in the questionnaire, and each participant independently evaluated the commits.

\begin{table}[!tp]
    \centering
    \caption{Scoring criterion for the human evaluation}
    \label{tab:scoring_criterion}
    \resizebox{\linewidth}{!}{
    \begin{tabular}{llp{0.7\textwidth}} 
        \toprule
        \textbf{Score} & \textbf{Definition} \\
        \midrule
        0 & Neither relevant in semantic nor having shared tokens. \\
        1 & Irrelevant in semantic but with some shared tokens. \\
        2 & Partially similar in semantic, but each contains exclusive information.\\
        3 & Highly similar but not identical in semantic. \\
        4 & Identical in semantic. \\
        \bottomrule
    \end{tabular}}
\end{table}

\subsubsection{Results}
For each technique, we assessed the quality of generated commit messages based on the average scores assigned by the six participants. 
Consistent with prior research \cite{DBLP:conf/icsm/Liu0K0BT18, Wang_tosem2021, Dong_icse2022}, commit messages scored 0 or 1 were categorized as low-quality, those scored 2 as medium-quality, and those scored 3 or 4 as high-quality. 
Table~\ref{tab:human_evaluation_results} summarizes the distribution of commit message quality across the techniques. 
Notably, 17\% of the commit messages generated by CONTENT were rated as high-quality by participants. 
CONTENT also produced the highest proportion of medium-quality messages and the lowest proportion of low-quality messages. 
Additionally, the average scores further highlighted CONTENT’s superior performance compared to NNGen, FIRA, and COME. 
To validate these findings, we conducted a Wilcoxon signed-rank test \cite{wilcoxon1992individual} to compare CONTENT’s scores against those of NNGen, FIRA, and COME. 
The p-value (< 0.05) indicated that the performance differences were statistically significant at the 95\% confidence level, confirming CONTENT’s superiority in generating high-quality commit messages.

\begin{table}[h]
    \centering
    \caption{Results of the human evaluation}
    \label{tab:human_evaluation_results}
    \resizebox{\linewidth}{!}{
    \begin{tabular}{lcccc}
        \toprule
        \textbf{Model} & \textbf{Low (\%)} & \textbf{Medium (\%)} & \textbf{High (\%)} & \textbf{Average Score} \\
        \midrule
        NNGen & 87.7 & 7.3 & 5.0 & 0.55 \\
        FIRA & 69.0 & 17.6 & 13.3 & 1.10 \\
        COME & 58.7 & 24.7 & 16.7 & 1.33 \\ \midrule
        CONTENT & 51.7 & 31.3 & 17.0 & 1.45 \\ 
        \bottomrule
    \end{tabular}}
    
\end{table}


\section{Threats and Limitations}
\subsection{Threats to Validity}

\textit{Internal validity concerns} including implementation errors and hyperparameter configurations – were mitigated by utilizing original authors' code and results where available, or by replicating experiments based on published descriptions to ensure consistency with prior findings. Additionally, human evaluation introduces validity threats due to participants' divergent scoring standards. To address this, we provided detailed evaluation criteria with concrete examples beforehand and recruited six participants (rather than three) to enhance reliability, as supported by \cite{DBLP:journals/tosem/WangXLHWG21}. Finally, commit message quality was assessed by averaging participant scores to minimize individual deviations.
\textit{External validity concerns}, related to dataset and metric selection, were mitigated by using the publicly available CoDiSum dataset and three established evaluation metrics (i.e., BLEU-Norm, METEOR, and ROUGE-L), further validated through human evaluations by experienced developers to confirm grammatical and semantic accuracy.

\subsection{Limitations}


We identify four primary limitations of CONTENT. 
First, if ChangeScribe cannot convert modified code into a condensation template—due to version mismatches—the method fails to generate a complete commit message. 
Although this issue did not occur with the enhanced ChangeScribe tool in our 87,602-sample dataset, it remains a practical concern. 
Second, CONTENT underperforms compared to retrieval-based approaches when the training data lacks diversity, particularly for inputs resembling rare examples. 
We suggest that leveraging more diverse training datasets could improve the generalization capabilities of pre-trained models, enabling them to handle diverse code changes effectively.
Third, existing commit message datasets—constructed from GitHub's top repositories—often exhibit weak semantic relevance between code changes and commit messages. Furthermore, while commit messages admit diverse formulations, current datasets contain only a single reference per instance, constraining output diversity.
Finally, we exclude closed-source LLMs (e.g., GPT-4) from our baseline, as prior research suggests they perform worse in commit message generation than learning-based and hybrid methods \cite{wu2024commit,li2024only}. 
However, we plan to explore their integration with our template in future work.





\section{Related Work}
\label{sec:relatedwork}
A growing body of work is proposed for commit message generation. They can be categorized into four types: template-based, retrieval-based, learning-based, and hybrid methods, described in what follows.

\textbf{Template-based methods} \cite{Buse_ase2010, Cortes-Coy_2014, Vasquez_icse2015, Shen_COMPSAC2016} generate commit messages by applying predefined templates to code changes. For example, Buse et al. \cite{Buse_ase2010} used symbolic execution to derive path predicates and generate messages using rule-based templates, while Cortés-Coy et al. \cite{Cortes-Coy_2014} developed stereotype-based templates to produce structured messages. However, these methods often lack flexibility for non-standard changes.
Yet, they fail in capturing the underlying motivation for a change, focusing primarily on what changed rather than why.

\textbf{Retrieval-based methods} \cite{Huang_esem2017, Liu_ase2018, Hoang_icse2020, Huang_jcst2020, shi_emnlp2022} generate messages by finding similar code changes and reusing their associated human-written messages. For instance, Liu et al. \cite{Liu_ase2018} used bag-of-words vectors to rank similar changes, while Huang et al. \cite{Huang_jcst2020} enhanced retrieval by incorporating syntax and semantics. Hoang et al. \cite{Hoang_icse2020} leveraged distributed representations of code changes to improve retrieval accuracy. However, these methods struggle when no similar changes exist \cite{tian_icse2022} and cannot generate new, contextually relevant messages.

\textbf{Learning-based methods} \cite{Loyola_acl2017, Jiang_ase2017, Xu_ijcai2019, Liu_msr2019, Dong_icse2022} model commit message generation as a \texttt{Seq2Seq} task. Jiang et al. \cite{Jiang_ase2017} treated code changes as input sequences, while Loyola et al. \cite{Loyola_acl2017} used an encoder-decoder model with attention \cite{Luong_emnlp2015} to improve message relevance. Liu et al. \cite{Liu_msr2019} adopted pointer-generator networks \cite{See_acl2017} to address vocabulary limitations, and others \cite{Xu_ijcai2019, Liu_tse2022} leveraged abstract syntax trees to model structural features. Dong et al. \cite{Dong_icse2022} represented code changes with fine-grained graphs using graph neural networks. Wang et al. \cite{Wang_ase2023} integrated commit-issue correlations into pre-trained models like CodeBERT \cite{Feng_emnlp20} and CodeT5 \cite{Wang_emnlp2021}. While these approaches use advanced models and fine-grained representations, our work focuses on textual representations for readability and extensibility, with potential for future integration with fine-grained methods.

\textbf{Hybrid approaches} \cite{He_ISSTA2023, Liu_tse2022, Wang_tosem2021} combine retrieval-based and learning-based methods. Shi et al.~\cite{shi_emnlp2022} integrated retrieval with generation models for better adaptability, while Wang et al.~\cite{Wang_tosem2021} addressed low-frequency vocabulary and exposure bias. He et al.~\cite{He_ISSTA2023} developed COME, combining fine-grained code change representations with retrieval and translation models. Liu et al. \cite{Liu_tse2022} used abstract syntax trees to inform both learning and retrieval modules. However, without a well-designed fusion mechanism, retrieval and learning components may interfere, reducing output quality \cite{He_ISSTA2023, Li_fse2024}.

Recently, LLMs like ChatGPT and GPT-4 have been applied to software engineering tasks, including code generation \cite{Fakhoury_tse2024, Mu_fse2024}, test case generation \cite{Xue_issta2024}, and traceability recovery \cite{Rodriguez_rew2023}. In CMG, Tao et al. \cite{Tao_tosem2024} introduced KADEL, a knowledge-aware denoising method, and compared it with ChatGPT, finding KADEL superior in automatic metrics but ChatGPT better in conciseness and expressiveness. Li et al. \cite{Li_fse2024} proposed OMG, a GPT-4-based method using the ReAct prompting strategy \cite{Yao_iclr2023} to integrate contextual information dynamically. In future work, we plan to explore integrating our approach with LLMs to further enhance CMG effectiveness.

\section{Conclusion and Future Work}
\label{sec:conclusion}

In this study, we introduce CONTENT, an innovative approach for generating high-quality commit messages. 
CONTENT first converts code changes into concise text templates incorporating summarized modifications, elicited comments, and key code identifiers. 
It then fine-tunes a CodeLlama-7B model to generate commit messages automatically.
By simplifying complex code changes into clearer, enriched descriptions, CONTENT effectively bridges the semantic gap between code modifications and commit messages.

Extensive evaluations on a benchmark dataset show that CONTENT outperforms six existing techniques across BLEU-Norm, ROUGE-L, and METEOR metrics. 
Ablation analysis confirms that its three key components each contribute positively to its performance. 
Moreover, experiments with different pre-trained models, including CodeBERT and CodeLlama-7B, demonstrate CONTENT’s robust generalization ability.
A human evaluation further confirmed its practical value from developers’ perspectives, 

For future work, we aim to enhance the semantic alignment between generated messages and code changes by incorporating developer intent, project conventions, and graph-based code dependency representations with advanced natural language processing techniques. 
We also plan to scale CONTENT for large projects and refine the code change summarization process.

\begin{acks} 
Feng Dai, Jiachen Li, Ziqi Wang and Yunqi Cui also contributed to this work.
This work is supported by the National Natural Science Foundation of China (No.62202219, No.62302210), the Jiangsu Provincial Key Research and Development Program (No.BE2021002-2), the Natural Science Foundation of Jiangsu Province (No.BK20241195), and the Innovation Project and Overseas Open Project of State Key Laboratory for Novel Software Technology (Nanjing University) (ZZKT2025A12, ZZKT2025B18, ZZKT2025B20, ZZKT2025B22, KFKT2025A17, KFKT2025A19, KFKT2025A20, KFKT2024A02, KFKT2024A13, KFKT2024A14, KFKT2023A09, KFKT2023A10).
\end{acks}

\bibliographystyle{ACM-Reference-Format}
\bibliography{manuscript}


\begin{thebibliography}{68}


\ifx \showCODEN    \undefined \def \showCODEN     #1{\unskip}     \fi
\ifx \showDOI      \undefined \def \showDOI       #1{#1}\fi
\ifx \showISBNx    \undefined \def \showISBNx     #1{\unskip}     \fi
\ifx \showISBNxiii \undefined \def \showISBNxiii  #1{\unskip}     \fi
\ifx \showISSN     \undefined \def \showISSN      #1{\unskip}     \fi
\ifx \showLCCN     \undefined \def \showLCCN      #1{\unskip}     \fi
\ifx \shownote     \undefined \def \shownote      #1{#1}          \fi
\ifx \showarticletitle \undefined \def \showarticletitle #1{#1}   \fi
\ifx \showURL      \undefined \def \showURL       {\relax}        \fi
\providecommand\bibfield[2]{#2}
\providecommand\bibinfo[2]{#2}
\providecommand\natexlab[1]{#1}
\providecommand\showeprint[2][]{arXiv:#2}

\bibitem[Git(2024)]%
        {GitHub}
 \bibinfo{year}{2024}\natexlab{}.
\newblock \bibinfo{booktitle}{\emph{GitHub}}.
\newblock
\urldef\tempurl%
\url{https://github.com/}
\showURL{%
\tempurl}


\bibitem[Sql(2024)]%
        {SqlConfig}
 \bibinfo{year}{2024}\natexlab{}.
\newblock \bibinfo{booktitle}{\emph{Spring SqlConfig Annotation}}.
\newblock
\urldef\tempurl%
\url{https://docs.spring.io/spring-framework/reference/testing/annotations/integration-spring/annotation-sqlconfig.html}
\showURL{%
\tempurl}


\bibitem[Tes(2024)]%
        {TestPropertySource}
 \bibinfo{year}{2024}\natexlab{}.
\newblock \bibinfo{booktitle}{\emph{Spring TestPropertySource Annotation}}.
\newblock
\urldef\tempurl%
\url{https://docs.spring.io/spring-framework/reference/testing/annotations/integration-spring/annotation-testpropertysource.html}
\showURL{%
\tempurl}


\bibitem[Ali et~al\mbox{.}(2015)]%
        {DBLP:journals/ese/AliSGA15}
\bibfield{author}{\bibinfo{person}{Nasir Ali}, \bibinfo{person}{Zohreh Sharafi}, \bibinfo{person}{Yann{-}Ga{\"{e}}l Gu{\'{e}}h{\'{e}}neuc}, {and} \bibinfo{person}{Giuliano Antoniol}.} \bibinfo{year}{2015}\natexlab{}.
\newblock \showarticletitle{An empirical study on the importance of source code entities for requirements traceability}.
\newblock \bibinfo{journal}{\emph{Empir. Softw. Eng.}} \bibinfo{volume}{20}, \bibinfo{number}{2} (\bibinfo{year}{2015}), \bibinfo{pages}{442--478}.
\newblock


\bibitem[Banerjee and Lavie(2005)]%
        {METEOR}
\bibfield{author}{\bibinfo{person}{Satanjeev Banerjee} {and} \bibinfo{person}{Alon Lavie}.} \bibinfo{year}{2005}\natexlab{}.
\newblock \showarticletitle{{METEOR:} An Automatic Metric for {MT} Evaluation with Improved Correlation with Human Judgments}. In \bibinfo{booktitle}{\emph{Proceedings of the Workshop on Intrinsic and Extrinsic Evaluation Measures for Machine Translation and/or Summarization@ACL 2005, Ann Arbor, Michigan, USA, June 29, 2005}}. \bibinfo{publisher}{Association for Computational Linguistics}, \bibinfo{pages}{65--72}.
\newblock


\bibitem[Buse and Weimer(2010)]%
        {Buse_ase2010}
\bibfield{author}{\bibinfo{person}{Raymond P.~L. Buse} {and} \bibinfo{person}{Westley Weimer}.} \bibinfo{year}{2010}\natexlab{}.
\newblock \showarticletitle{Automatically documenting program changes}. In \bibinfo{booktitle}{\emph{{ASE} 2010, 25th {IEEE/ACM} International Conference on Automated Software Engineering, Antwerp, Belgium, September 20-24, 2010}}. \bibinfo{publisher}{{ACM}}, \bibinfo{pages}{33--42}.
\newblock


\bibitem[Cortes{-}Coy et~al\mbox{.}(2014a)]%
        {Cortes-Coy_2014}
\bibfield{author}{\bibinfo{person}{Luis~Fernando Cortes{-}Coy}, \bibinfo{person}{Mario~Linares V{\'{a}}squez}, \bibinfo{person}{Jairo Aponte}, {and} \bibinfo{person}{Denys Poshyvanyk}.} \bibinfo{year}{2014}\natexlab{a}.
\newblock \showarticletitle{On Automatically Generating Commit Messages via Summarization of Source Code Changes}. In \bibinfo{booktitle}{\emph{14th {IEEE} International Working Conference on Source Code Analysis and Manipulation, {SCAM} 2014, Victoria, BC, Canada, September 28-29, 2014}}. \bibinfo{publisher}{{IEEE} Computer Society}, \bibinfo{pages}{275--284}.
\newblock


\bibitem[Cortes{-}Coy et~al\mbox{.}(2014b)]%
        {ChangeScribe}
\bibfield{author}{\bibinfo{person}{Luis~Fernando Cortes{-}Coy}, \bibinfo{person}{Mario~Linares V{\'{a}}squez}, \bibinfo{person}{Jairo Aponte}, {and} \bibinfo{person}{Denys Poshyvanyk}.} \bibinfo{year}{2014}\natexlab{b}.
\newblock \showarticletitle{On Automatically Generating Commit Messages via Summarization of Source Code Changes}. In \bibinfo{booktitle}{\emph{14th {IEEE} International Working Conference on Source Code Analysis and Manipulation, {SCAM} 2014, Victoria, BC, Canada, September 28-29, 2014}}. \bibinfo{publisher}{{IEEE} Computer Society}, \bibinfo{pages}{275--284}.
\newblock


\bibitem[Dong et~al\mbox{.}(2023)]%
        {DBLP:conf/icse/DongLHT23}
\bibfield{author}{\bibinfo{person}{Jinhao Dong}, \bibinfo{person}{Yiling Lou}, \bibinfo{person}{Dan Hao}, {and} \bibinfo{person}{Lin Tan}.} \bibinfo{year}{2023}\natexlab{}.
\newblock \showarticletitle{Revisiting Learning-based Commit Message Generation}. In \bibinfo{booktitle}{\emph{45th {IEEE/ACM} International Conference on Software Engineering, {ICSE} 2023, Melbourne, Australia, May 14-20, 2023}}. \bibinfo{publisher}{{IEEE}}, \bibinfo{pages}{794--805}.
\newblock


\bibitem[Dong et~al\mbox{.}(2022)]%
        {Dong_icse2022}
\bibfield{author}{\bibinfo{person}{Jinhao Dong}, \bibinfo{person}{Yiling Lou}, \bibinfo{person}{Qihao Zhu}, \bibinfo{person}{Zeyu Sun}, \bibinfo{person}{Zhilin Li}, \bibinfo{person}{Wenjie Zhang}, {and} \bibinfo{person}{Dan Hao}.} \bibinfo{year}{2022}\natexlab{}.
\newblock \showarticletitle{{FIRA:} Fine-Grained Graph-Based Code Change Representation for Automated Commit Message Generation}. In \bibinfo{booktitle}{\emph{44th {IEEE/ACM} 44th International Conference on Software Engineering, {ICSE} 2022, Pittsburgh, PA, USA, May 25-27, 2022}}. \bibinfo{publisher}{{ACM}}, \bibinfo{pages}{970--981}.
\newblock


\bibitem[Dyer et~al\mbox{.}(2013)]%
        {DBLP:conf/icse/0001NRN13}
\bibfield{author}{\bibinfo{person}{Robert Dyer}, \bibinfo{person}{Hoan~Anh Nguyen}, \bibinfo{person}{Hridesh Rajan}, {and} \bibinfo{person}{Tien~N. Nguyen}.} \bibinfo{year}{2013}\natexlab{}.
\newblock \showarticletitle{Boa: a language and infrastructure for analyzing ultra-large-scale software repositories}. In \bibinfo{booktitle}{\emph{35th International Conference on Software Engineering, {ICSE} '13, San Francisco, CA, USA, May 18-26, 2013}}. \bibinfo{publisher}{{IEEE} Computer Society}, \bibinfo{pages}{422--431}.
\newblock


\bibitem[Fakhoury et~al\mbox{.}(2024)]%
        {Fakhoury_tse2024}
\bibfield{author}{\bibinfo{person}{Sarah Fakhoury}, \bibinfo{person}{Aaditya Naik}, \bibinfo{person}{Georgios Sakkas}, \bibinfo{person}{Saikat Chakraborty}, {and} \bibinfo{person}{Shuvendu~K. Lahiri}.} \bibinfo{year}{2024}\natexlab{}.
\newblock \showarticletitle{LLM-Based Test-Driven Interactive Code Generation: User Study and Empirical Evaluation}.
\newblock \bibinfo{journal}{\emph{{IEEE} Trans. Software Eng.}} \bibinfo{volume}{50}, \bibinfo{number}{9} (\bibinfo{year}{2024}), \bibinfo{pages}{2254--2268}.
\newblock


\bibitem[Feng et~al\mbox{.}(2020)]%
        {Feng_emnlp20}
\bibfield{author}{\bibinfo{person}{Zhangyin Feng}, \bibinfo{person}{Daya Guo}, \bibinfo{person}{Duyu Tang}, \bibinfo{person}{Nan Duan}, \bibinfo{person}{Xiaocheng Feng}, \bibinfo{person}{Ming Gong}, \bibinfo{person}{Linjun Shou}, \bibinfo{person}{Bing Qin}, \bibinfo{person}{Ting Liu}, \bibinfo{person}{Daxin Jiang}, {and} \bibinfo{person}{Ming Zhou}.} \bibinfo{year}{2020}\natexlab{}.
\newblock \showarticletitle{CodeBERT: {A} Pre-Trained Model for Programming and Natural Languages}. In \bibinfo{booktitle}{\emph{Findings of the Association for Computational Linguistics: {EMNLP} 2020, Online Event, 16-20 November 2020}} \emph{(\bibinfo{series}{Findings of {ACL}}, Vol.~\bibinfo{volume}{{EMNLP} 2020})}. \bibinfo{publisher}{Association for Computational Linguistics}, \bibinfo{pages}{1536--1547}.
\newblock


\bibitem[He et~al\mbox{.}(2023a)]%
        {He_ISSTA2023}
\bibfield{author}{\bibinfo{person}{Yichen He}, \bibinfo{person}{Liran Wang}, \bibinfo{person}{Kaiyi Wang}, \bibinfo{person}{Yupeng Zhang}, \bibinfo{person}{Hang Zhang}, {and} \bibinfo{person}{Zhoujun Li}.} \bibinfo{year}{2023}\natexlab{a}.
\newblock \showarticletitle{{COME:} Commit Message Generation with Modification Embedding}. In \bibinfo{booktitle}{\emph{Proceedings of the 32nd {ACM} {SIGSOFT} International Symposium on Software Testing and Analysis, {ISSTA} 2023, Seattle, WA, USA, July 17-21, 2023}}. \bibinfo{publisher}{{ACM}}, \bibinfo{pages}{792--803}.
\newblock


\bibitem[He et~al\mbox{.}(2023b)]%
        {DBLP:conf/issta/HeWWZZ023}
\bibfield{author}{\bibinfo{person}{Yichen He}, \bibinfo{person}{Liran Wang}, \bibinfo{person}{Kaiyi Wang}, \bibinfo{person}{Yupeng Zhang}, \bibinfo{person}{Hang Zhang}, {and} \bibinfo{person}{Zhoujun Li}.} \bibinfo{year}{2023}\natexlab{b}.
\newblock \showarticletitle{{COME:} Commit Message Generation with Modification Embedding}. In \bibinfo{booktitle}{\emph{Proceedings of the 32nd {ACM} {SIGSOFT} International Symposium on Software Testing and Analysis, {ISSTA} 2023, Seattle, WA, USA, July 17-21, 2023}}. \bibinfo{publisher}{{ACM}}, \bibinfo{pages}{792--803}.
\newblock


\bibitem[Hoang et~al\mbox{.}(2020)]%
        {Hoang_icse2020}
\bibfield{author}{\bibinfo{person}{Thong Hoang}, \bibinfo{person}{Hong~Jin Kang}, \bibinfo{person}{David Lo}, {and} \bibinfo{person}{Julia Lawall}.} \bibinfo{year}{2020}\natexlab{}.
\newblock \showarticletitle{CC2Vec: distributed representations of code changes}. In \bibinfo{booktitle}{\emph{{ICSE} '20: 42nd International Conference on Software Engineering, Seoul, South Korea, 27 June - 19 July, 2020}}. \bibinfo{publisher}{{ACM}}, \bibinfo{pages}{518--529}.
\newblock


\bibitem[Hu et~al\mbox{.}(2024)]%
        {Hu_ase2024}
\bibfield{author}{\bibinfo{person}{Chao Hu}, \bibinfo{person}{Yitian Chai}, \bibinfo{person}{Hao Zhou}, \bibinfo{person}{Fandong Meng}, \bibinfo{person}{Jie Zhou}, {and} \bibinfo{person}{Xiaodong Gu}.} \bibinfo{year}{2024}\natexlab{}.
\newblock \showarticletitle{How Effectively Do Code Language Models Understand Poor-Readability Code?}. In \bibinfo{booktitle}{\emph{Proceedings of the 39th {IEEE/ACM} International Conference on Automated Software Engineering, {ASE} 2024, Sacramento, CA, USA, October 27 - November 1, 2024}}. \bibinfo{publisher}{{ACM}}, \bibinfo{pages}{795--806}.
\newblock


\bibitem[Huang et~al\mbox{.}(2020)]%
        {Huang_jcst2020}
\bibfield{author}{\bibinfo{person}{Yuan Huang}, \bibinfo{person}{Nan Jia}, \bibinfo{person}{Haojie Zhou}, \bibinfo{person}{Xiangping Chen}, \bibinfo{person}{Zibin Zheng}, {and} \bibinfo{person}{Mingdong Tang}.} \bibinfo{year}{2020}\natexlab{}.
\newblock \showarticletitle{Learning Human-Written Commit Messages to Document Code Changes}.
\newblock \bibinfo{journal}{\emph{J. Comput. Sci. Technol.}} \bibinfo{volume}{35}, \bibinfo{number}{6} (\bibinfo{year}{2020}), \bibinfo{pages}{1258--1277}.
\newblock


\bibitem[Huang et~al\mbox{.}(2017)]%
        {Huang_esem2017}
\bibfield{author}{\bibinfo{person}{Yuan Huang}, \bibinfo{person}{Qiaoyang Zheng}, \bibinfo{person}{Xiangping Chen}, \bibinfo{person}{Yingfei Xiong}, \bibinfo{person}{Zhiyong Liu}, {and} \bibinfo{person}{Xiaonan Luo}.} \bibinfo{year}{2017}\natexlab{}.
\newblock \showarticletitle{Mining Version Control System for Automatically Generating Commit Comment}. In \bibinfo{booktitle}{\emph{2017 {ACM/IEEE} International Symposium on Empirical Software Engineering and Measurement, {ESEM} 2017, Toronto, ON, Canada, November 9-10, 2017}}. \bibinfo{publisher}{{IEEE} Computer Society}, \bibinfo{pages}{414--423}.
\newblock


\bibitem[Jiang et~al\mbox{.}(2017a)]%
        {DBLP:conf/kbse/JiangAM17}
\bibfield{author}{\bibinfo{person}{Siyuan Jiang}, \bibinfo{person}{Ameer Armaly}, {and} \bibinfo{person}{Collin McMillan}.} \bibinfo{year}{2017}\natexlab{a}.
\newblock \showarticletitle{Automatically generating commit messages from diffs using neural machine translation}. In \bibinfo{booktitle}{\emph{Proceedings of the 32nd {IEEE/ACM} International Conference on Automated Software Engineering, {ASE} 2017, Urbana, IL, USA, October 30 - November 03, 2017}}. \bibinfo{publisher}{{IEEE} Computer Society}, \bibinfo{pages}{135--146}.
\newblock


\bibitem[Jiang et~al\mbox{.}(2017b)]%
        {Jiang_ase2017}
\bibfield{author}{\bibinfo{person}{Siyuan Jiang}, \bibinfo{person}{Ameer Armaly}, {and} \bibinfo{person}{Collin McMillan}.} \bibinfo{year}{2017}\natexlab{b}.
\newblock \showarticletitle{Automatically generating commit messages from diffs using neural machine translation}. In \bibinfo{booktitle}{\emph{Proceedings of the 32nd {IEEE/ACM} International Conference on Automated Software Engineering, {ASE} 2017, Urbana, IL, USA, October 30 - November 03, 2017}}. \bibinfo{publisher}{{IEEE} Computer Society}, \bibinfo{pages}{135--146}.
\newblock


\bibitem[Kajko{-}Mattsson(2005)]%
        {DBLP:journals/ese/Kajko-Mattsson05}
\bibfield{author}{\bibinfo{person}{Mira Kajko{-}Mattsson}.} \bibinfo{year}{2005}\natexlab{}.
\newblock \showarticletitle{A Survey of Documentation Practice within Corrective Maintenance}.
\newblock \bibinfo{journal}{\emph{Empir. Softw. Eng.}} \bibinfo{volume}{10}, \bibinfo{number}{1} (\bibinfo{year}{2005}), \bibinfo{pages}{31--55}.
\newblock


\bibitem[Lebedev(2017)]%
        {javalang}
\bibfield{author}{\bibinfo{person}{Sergei Lebedev}.} \bibinfo{year}{2017}\natexlab{}.
\newblock \bibinfo{title}{javalang: Pure Python Java Parser}.
\newblock
\newblock
\urldef\tempurl%
\url{https://github.com/c2nes/javalang}
\showURL{%
\tempurl}


\bibitem[Li et~al\mbox{.}(2024a)]%
        {Li_fse2024}
\bibfield{author}{\bibinfo{person}{Jiawei Li}, \bibinfo{person}{David Farag{\'{o}}}, \bibinfo{person}{Christian Petrov}, {and} \bibinfo{person}{Iftekhar Ahmed}.} \bibinfo{year}{2024}\natexlab{a}.
\newblock \showarticletitle{Only diff Is Not Enough: Generating Commit Messages Leveraging Reasoning and Action of Large Language Model}.
\newblock \bibinfo{journal}{\emph{Proc. {ACM} Softw. Eng.}} \bibinfo{volume}{1}, \bibinfo{number}{{FSE}} (\bibinfo{year}{2024}), \bibinfo{pages}{745--766}.
\newblock


\bibitem[Li et~al\mbox{.}(2024b)]%
        {li2024only}
\bibfield{author}{\bibinfo{person}{Jiawei Li}, \bibinfo{person}{David Farag{\'o}}, \bibinfo{person}{Christian Petrov}, {and} \bibinfo{person}{Iftekhar Ahmed}.} \bibinfo{year}{2024}\natexlab{b}.
\newblock \showarticletitle{Only diff is not enough: Generating commit messages leveraging reasoning and action of large language model}.
\newblock \bibinfo{journal}{\emph{Proceedings of the ACM on Software Engineering}} \bibinfo{volume}{1}, \bibinfo{number}{FSE} (\bibinfo{year}{2024}), \bibinfo{pages}{745--766}.
\newblock


\bibitem[Li et~al\mbox{.}(2018)]%
        {conf/iwpc/LiNJWHW18}
\bibfield{author}{\bibinfo{person}{Shanshan Li}, \bibinfo{person}{Xu Niu}, \bibinfo{person}{Zhouyang Jia}, \bibinfo{person}{Ji Wang}, \bibinfo{person}{Haochen He}, {and} \bibinfo{person}{Teng Wang}.} \bibinfo{year}{2018}\natexlab{}.
\newblock \showarticletitle{Logtracker: learning log revision behaviors proactively from software evolution history}. In \bibinfo{booktitle}{\emph{Proceedings of the 26th Conference on Program Comprehension, {ICPC} 2018, Gothenburg, Sweden, May 27-28, 2018}}. \bibinfo{publisher}{{ACM}}, \bibinfo{pages}{178--188}.
\newblock


\bibitem[Liu et~al\mbox{.}(2018a)]%
        {DBLP:conf/icsm/Liu0K0BT18}
\bibfield{author}{\bibinfo{person}{Kui Liu}, \bibinfo{person}{Dongsun Kim}, \bibinfo{person}{Anil Koyuncu}, \bibinfo{person}{Li Li}, \bibinfo{person}{Tegawend{\'{e}}~F. Bissyand{\'{e}}}, {and} \bibinfo{person}{Yves~Le Traon}.} \bibinfo{year}{2018}\natexlab{a}.
\newblock \showarticletitle{A Closer Look at Real-World Patches}. In \bibinfo{booktitle}{\emph{2018 {IEEE} International Conference on Software Maintenance and Evolution, {ICSME} 2018, Madrid, Spain, September 23-29, 2018}}. \bibinfo{publisher}{{IEEE} Computer Society}, \bibinfo{pages}{275--286}.
\newblock


\bibitem[Liu et~al\mbox{.}(2019)]%
        {Liu_msr2019}
\bibfield{author}{\bibinfo{person}{Qin Liu}, \bibinfo{person}{Zihe Liu}, \bibinfo{person}{Hongming Zhu}, \bibinfo{person}{Hongfei Fan}, \bibinfo{person}{Bowen Du}, {and} \bibinfo{person}{Yu Qian}.} \bibinfo{year}{2019}\natexlab{}.
\newblock \showarticletitle{Generating commit messages from diffs using pointer-generator network}. In \bibinfo{booktitle}{\emph{Proceedings of the 16th International Conference on Mining Software Repositories, {MSR} 2019, 26-27 May 2019, Montreal, Canada}}. \bibinfo{publisher}{{IEEE} / {ACM}}, \bibinfo{pages}{299--309}.
\newblock


\bibitem[Liu et~al\mbox{.}(2022)]%
        {Liu_tse2022}
\bibfield{author}{\bibinfo{person}{Shangqing Liu}, \bibinfo{person}{Cuiyun Gao}, \bibinfo{person}{Sen Chen}, \bibinfo{person}{Lun~Yiu Nie}, {and} \bibinfo{person}{Yang Liu}.} \bibinfo{year}{2022}\natexlab{}.
\newblock \showarticletitle{{ATOM:} Commit Message Generation Based on Abstract Syntax Tree and Hybrid Ranking}.
\newblock \bibinfo{journal}{\emph{{IEEE} Trans. Software Eng.}} \bibinfo{volume}{48}, \bibinfo{number}{5} (\bibinfo{year}{2022}), \bibinfo{pages}{1800--1817}.
\newblock


\bibitem[Liu(2019)]%
        {liu2019roberta}
\bibfield{author}{\bibinfo{person}{Yinhan Liu}.} \bibinfo{year}{2019}\natexlab{}.
\newblock \showarticletitle{Roberta: A robustly optimized bert pretraining approach}.
\newblock \bibinfo{journal}{\emph{arXiv preprint arXiv:1907.11692}}  \bibinfo{volume}{364} (\bibinfo{year}{2019}).
\newblock


\bibitem[Liu et~al\mbox{.}(2018b)]%
        {Liu_ase2018}
\bibfield{author}{\bibinfo{person}{Zhongxin Liu}, \bibinfo{person}{Xin Xia}, \bibinfo{person}{Ahmed~E. Hassan}, \bibinfo{person}{David Lo}, \bibinfo{person}{Zhenchang Xing}, {and} \bibinfo{person}{Xinyu Wang}.} \bibinfo{year}{2018}\natexlab{b}.
\newblock \showarticletitle{Neural-machine-translation-based commit message generation: how far are we?}. In \bibinfo{booktitle}{\emph{Proceedings of the 33rd {ACM/IEEE} International Conference on Automated Software Engineering, {ASE} 2018, Montpellier, France, September 3-7, 2018}}. \bibinfo{publisher}{{ACM}}, \bibinfo{pages}{373--384}.
\newblock


\bibitem[Liu et~al\mbox{.}(2018c)]%
        {liu2018neural}
\bibfield{author}{\bibinfo{person}{Zhongxin Liu}, \bibinfo{person}{Xin Xia}, \bibinfo{person}{Ahmed~E Hassan}, \bibinfo{person}{David Lo}, \bibinfo{person}{Zhenchang Xing}, {and} \bibinfo{person}{Xinyu Wang}.} \bibinfo{year}{2018}\natexlab{c}.
\newblock \showarticletitle{Neural-machine-translation-based commit message generation: how far are we?}. In \bibinfo{booktitle}{\emph{Proceedings of the 33rd ACM/IEEE International Conference on Automated Software Engineering}}. \bibinfo{pages}{373--384}.
\newblock


\bibitem[Loyola et~al\mbox{.}(2017)]%
        {Loyola_acl2017}
\bibfield{author}{\bibinfo{person}{Pablo Loyola}, \bibinfo{person}{Edison Marrese{-}Taylor}, {and} \bibinfo{person}{Yutaka Matsuo}.} \bibinfo{year}{2017}\natexlab{}.
\newblock \showarticletitle{A Neural Architecture for Generating Natural Language Descriptions from Source Code Changes}. In \bibinfo{booktitle}{\emph{Proceedings of the 55th Annual Meeting of the Association for Computational Linguistics, {ACL} 2017, Vancouver, Canada, July 30 - August 4, Volume 2: Short Papers}}. \bibinfo{publisher}{Association for Computational Linguistics}, \bibinfo{pages}{287--292}.
\newblock


\bibitem[Luong et~al\mbox{.}(2015)]%
        {Luong_emnlp2015}
\bibfield{author}{\bibinfo{person}{Thang Luong}, \bibinfo{person}{Hieu Pham}, {and} \bibinfo{person}{Christopher~D. Manning}.} \bibinfo{year}{2015}\natexlab{}.
\newblock \showarticletitle{Effective Approaches to Attention-based Neural Machine Translation}. In \bibinfo{booktitle}{\emph{Proceedings of the 2015 Conference on Empirical Methods in Natural Language Processing, {EMNLP} 2015, Lisbon, Portugal, September 17-21, 2015}}. \bibinfo{publisher}{The Association for Computational Linguistics}, \bibinfo{pages}{1412--1421}.
\newblock


\bibitem[Ma et~al\mbox{.}(2024)]%
        {Ma_tosem2024}
\bibfield{author}{\bibinfo{person}{Wei Ma}, \bibinfo{person}{Shangqing Liu}, \bibinfo{person}{Mengjie Zhao}, \bibinfo{person}{Xiaofei Xie}, \bibinfo{person}{Wenhan Wang}, \bibinfo{person}{Qiang Hu}, \bibinfo{person}{Jie Zhang}, {and} \bibinfo{person}{Yang Liu}.} \bibinfo{year}{2024}\natexlab{}.
\newblock \showarticletitle{Unveiling Code Pre-Trained Models: Investigating Syntax and Semantics Capacities}.
\newblock \bibinfo{journal}{\emph{{ACM} Trans. Softw. Eng. Methodol.}} \bibinfo{volume}{33}, \bibinfo{number}{7} (\bibinfo{year}{2024}), \bibinfo{pages}{169:1--169:29}.
\newblock


\bibitem[Maalej and Happel(2010)]%
        {DBLP:conf/msr/MaalejH10}
\bibfield{author}{\bibinfo{person}{Walid Maalej} {and} \bibinfo{person}{Hans{-}J{\"{o}}rg Happel}.} \bibinfo{year}{2010}\natexlab{}.
\newblock \showarticletitle{Can development work describe itself?}. In \bibinfo{booktitle}{\emph{Proceedings of the 7th International Working Conference on Mining Software Repositories, {MSR} 2010 (Co-located with ICSE), Cape Town, South Africa, May 2-3, 2010, Proceedings}}. \bibinfo{publisher}{{IEEE} Computer Society}, \bibinfo{pages}{191--200}.
\newblock


\bibitem[Mesnard et~al\mbox{.}(2024)]%
        {DBLP:journals/corr/abs-2403-08295}
\bibfield{author}{\bibinfo{person}{Thomas Mesnard}, \bibinfo{person}{Cassidy Hardin}, \bibinfo{person}{Robert Dadashi}, \bibinfo{person}{Surya Bhupatiraju}, \bibinfo{person}{Shreya Pathak}, \bibinfo{person}{Laurent Sifre}, \bibinfo{person}{Morgane Rivi{\`{e}}re}, \bibinfo{person}{Mihir~Sanjay Kale}, \bibinfo{person}{Juliette Love}, \bibinfo{person}{Pouya Tafti}, \bibinfo{person}{L{\'{e}}onard Hussenot}, \bibinfo{person}{Aakanksha Chowdhery}, \bibinfo{person}{Adam Roberts}, \bibinfo{person}{Aditya Barua}, \bibinfo{person}{Alex Botev}, \bibinfo{person}{Alex Castro{-}Ros}, {and} \bibinfo{person}{et al.}} \bibinfo{year}{2024}\natexlab{}.
\newblock \showarticletitle{Gemma: Open Models Based on Gemini Research and Technology}.
\newblock \bibinfo{journal}{\emph{CoRR}}  \bibinfo{volume}{abs/2403.08295} (\bibinfo{year}{2024}).
\newblock
\showeprint[arXiv]{2403.08295}


\bibitem[Mu et~al\mbox{.}(2024)]%
        {Mu_fse2024}
\bibfield{author}{\bibinfo{person}{Fangwen Mu}, \bibinfo{person}{Lin Shi}, \bibinfo{person}{Song Wang}, \bibinfo{person}{Zhuohao Yu}, \bibinfo{person}{Binquan Zhang}, \bibinfo{person}{Chenxue Wang}, \bibinfo{person}{Shichao Liu}, {and} \bibinfo{person}{Qing Wang}.} \bibinfo{year}{2024}\natexlab{}.
\newblock \showarticletitle{ClarifyGPT: {A} Framework for Enhancing LLM-Based Code Generation via Requirements Clarification}.
\newblock \bibinfo{journal}{\emph{Proc. {ACM} Softw. Eng.}} \bibinfo{volume}{1}, \bibinfo{number}{{FSE}} (\bibinfo{year}{2024}), \bibinfo{pages}{2332--2354}.
\newblock


\bibitem[Nie et~al\mbox{.}(2021a)]%
        {CoreGen}
\bibfield{author}{\bibinfo{person}{Lun~Yiu Nie}, \bibinfo{person}{Cuiyun Gao}, \bibinfo{person}{Zhicong Zhong}, \bibinfo{person}{Wai Lam}, \bibinfo{person}{Yang Liu}, {and} \bibinfo{person}{Zenglin Xu}.} \bibinfo{year}{2021}\natexlab{a}.
\newblock \showarticletitle{CoreGen: Contextualized Code Representation Learning for Commit Message Generation}.
\newblock \bibinfo{journal}{\emph{Neurocomputing}}  \bibinfo{volume}{459} (\bibinfo{year}{2021}), \bibinfo{pages}{97--107}.
\newblock


\bibitem[Nie et~al\mbox{.}(2021b)]%
        {DBLP:journals/ijon/NieGZLLX21}
\bibfield{author}{\bibinfo{person}{Lun~Yiu Nie}, \bibinfo{person}{Cuiyun Gao}, \bibinfo{person}{Zhicong Zhong}, \bibinfo{person}{Wai Lam}, \bibinfo{person}{Yang Liu}, {and} \bibinfo{person}{Zenglin Xu}.} \bibinfo{year}{2021}\natexlab{b}.
\newblock \showarticletitle{CoreGen: Contextualized Code Representation Learning for Commit Message Generation}.
\newblock \bibinfo{journal}{\emph{Neurocomputing}}  \bibinfo{volume}{459} (\bibinfo{year}{2021}), \bibinfo{pages}{97--107}.
\newblock


\bibitem[Papineni et~al\mbox{.}(2002)]%
        {acl/PapineniRWZ02}
\bibfield{author}{\bibinfo{person}{Kishore Papineni}, \bibinfo{person}{Salim Roukos}, \bibinfo{person}{Todd Ward}, {and} \bibinfo{person}{Wei{-}Jing Zhu}.} \bibinfo{year}{2002}\natexlab{}.
\newblock \showarticletitle{Bleu: a Method for Automatic Evaluation of Machine Translation}. In \bibinfo{booktitle}{\emph{Proceedings of the 40th Annual Meeting of the Association for Computational Linguistics, July 6-12, 2002, Philadelphia, PA, {USA}}}. \bibinfo{publisher}{{ACL}}, \bibinfo{pages}{311--318}.
\newblock


\bibitem[Qu et~al\mbox{.}(2024)]%
        {Qu_ASE2024}
\bibfield{author}{\bibinfo{person}{Muzi Qu}, \bibinfo{person}{Jie Liu}, \bibinfo{person}{Liangyi Kang}, \bibinfo{person}{Shuai Wang}, \bibinfo{person}{Dan Ye}, {and} \bibinfo{person}{Tao Huang}.} \bibinfo{year}{2024}\natexlab{}.
\newblock \showarticletitle{Dynamic Scoring Code Token Tree: {A} Novel Decoding Strategy for Generating High-Performance Code}. In \bibinfo{booktitle}{\emph{Proceedings of the 39th {IEEE/ACM} International Conference on Automated Software Engineering, {ASE} 2024, Sacramento, CA, USA, October 27 - November 1, 2024}}. \bibinfo{publisher}{{ACM}}, \bibinfo{pages}{1308--1318}.
\newblock


\bibitem[Rafailov et~al\mbox{.}(2023)]%
        {DBLP:conf/nips/RafailovSMMEF23}
\bibfield{author}{\bibinfo{person}{Rafael Rafailov}, \bibinfo{person}{Archit Sharma}, \bibinfo{person}{Eric Mitchell}, \bibinfo{person}{Christopher~D. Manning}, \bibinfo{person}{Stefano Ermon}, {and} \bibinfo{person}{Chelsea Finn}.} \bibinfo{year}{2023}\natexlab{}.
\newblock \showarticletitle{Direct Preference Optimization: Your Language Model is Secretly a Reward Model}. In \bibinfo{booktitle}{\emph{Advances in Neural Information Processing Systems 36: Annual Conference on Neural Information Processing Systems 2023, NeurIPS 2023, New Orleans, LA, USA, December 10 - 16, 2023}}.
\newblock


\bibitem[Rodriguez et~al\mbox{.}(2023)]%
        {Rodriguez_rew2023}
\bibfield{author}{\bibinfo{person}{Alberto~D. Rodriguez}, \bibinfo{person}{Katherine~R. Dearstyne}, {and} \bibinfo{person}{Jane Cleland{-}Huang}.} \bibinfo{year}{2023}\natexlab{}.
\newblock \showarticletitle{Prompts Matter: Insights and Strategies for Prompt Engineering in Automated Software Traceability}. In \bibinfo{booktitle}{\emph{31st {IEEE} International Requirements Engineering Conference, {RE} 2023 - Workshops, Hannover, Germany, September 4-5, 2023}}. \bibinfo{publisher}{{IEEE}}, \bibinfo{pages}{455--464}.
\newblock


\bibitem[Roziere et~al\mbox{.}(2023)]%
        {CodeLlama}
\bibfield{author}{\bibinfo{person}{Baptiste Roziere}, \bibinfo{person}{Jonas Gehring}, \bibinfo{person}{Fabian Gloeckle}, \bibinfo{person}{Sten Sootla}, \bibinfo{person}{Itai Gat}, \bibinfo{person}{Xiaoqing~Ellen Tan}, \bibinfo{person}{Yossi Adi}, \bibinfo{person}{Jingyu Liu}, \bibinfo{person}{Romain Sauvestre}, \bibinfo{person}{Tal Remez}, {et~al\mbox{.}}} \bibinfo{year}{2023}\natexlab{}.
\newblock \showarticletitle{Code llama: Open foundation models for code}.
\newblock \bibinfo{journal}{\emph{arXiv preprint arXiv:2308.12950}} (\bibinfo{year}{2023}).
\newblock


\bibitem[See et~al\mbox{.}(2017)]%
        {See_acl2017}
\bibfield{author}{\bibinfo{person}{Abigail See}, \bibinfo{person}{Peter~J. Liu}, {and} \bibinfo{person}{Christopher~D. Manning}.} \bibinfo{year}{2017}\natexlab{}.
\newblock \showarticletitle{Get To The Point: Summarization with Pointer-Generator Networks}. In \bibinfo{booktitle}{\emph{Proceedings of the 55th Annual Meeting of the Association for Computational Linguistics, {ACL} 2017, Vancouver, Canada, July 30 - August 4, Volume 1: Long Papers}}. \bibinfo{publisher}{Association for Computational Linguistics}, \bibinfo{pages}{1073--1083}.
\newblock


\bibitem[Shang et~al\mbox{.}(2024)]%
        {Shang_icsme2024}
\bibfield{author}{\bibinfo{person}{Xiuwei Shang}, \bibinfo{person}{Shaoyin Cheng}, \bibinfo{person}{Guoqiang Chen}, \bibinfo{person}{Yanming Zhang}, \bibinfo{person}{Li Hu}, \bibinfo{person}{Xiao Yu}, \bibinfo{person}{Gangyang Li}, \bibinfo{person}{Weiming Zhang}, {and} \bibinfo{person}{Nenghai Yu}.} \bibinfo{year}{2024}\natexlab{}.
\newblock \showarticletitle{How Far Have We Gone in Binary Code Understanding Using Large Language Models}. In \bibinfo{booktitle}{\emph{{IEEE} International Conference on Software Maintenance and Evolution, {ICSME} 2024, Flagstaff, AZ, USA, October 6-11, 2024}}. \bibinfo{publisher}{{IEEE}}, \bibinfo{pages}{1--12}.
\newblock


\bibitem[Shen et~al\mbox{.}(2016)]%
        {Shen_COMPSAC2016}
\bibfield{author}{\bibinfo{person}{Jinfeng Shen}, \bibinfo{person}{Xiaobing Sun}, \bibinfo{person}{Bin Li}, \bibinfo{person}{Hui Yang}, {and} \bibinfo{person}{Jiajun Hu}.} \bibinfo{year}{2016}\natexlab{}.
\newblock \showarticletitle{On Automatic Summarization of What and Why Information in Source Code Changes}. In \bibinfo{booktitle}{\emph{40th {IEEE} Annual Computer Software and Applications Conference, {COMPSAC} 2016, Atlanta, GA, USA, June 10-14, 2016}}. \bibinfo{publisher}{{IEEE} Computer Society}, \bibinfo{pages}{103--112}.
\newblock


\bibitem[Shi et~al\mbox{.}(2022)]%
        {shi_emnlp2022}
\bibfield{author}{\bibinfo{person}{Ensheng Shi}, \bibinfo{person}{Yanlin Wang}, \bibinfo{person}{Wei Tao}, \bibinfo{person}{Lun Du}, \bibinfo{person}{Hongyu Zhang}, \bibinfo{person}{Shi Han}, \bibinfo{person}{Dongmei Zhang}, {and} \bibinfo{person}{Hongbin Sun}.} \bibinfo{year}{2022}\natexlab{}.
\newblock \showarticletitle{{RACE:} Retrieval-augmented Commit Message Generation}. In \bibinfo{booktitle}{\emph{Proceedings of the 2022 Conference on Empirical Methods in Natural Language Processing, {EMNLP} 2022, Abu Dhabi, United Arab Emirates, December 7-11, 2022}}. \bibinfo{publisher}{Association for Computational Linguistics}, \bibinfo{pages}{5520--5530}.
\newblock


\bibitem[Tao et~al\mbox{.}(2021)]%
        {icsm/TaoWSDH0ZZ21}
\bibfield{author}{\bibinfo{person}{Wei Tao}, \bibinfo{person}{Yanlin Wang}, \bibinfo{person}{Ensheng Shi}, \bibinfo{person}{Lun Du}, \bibinfo{person}{Shi Han}, \bibinfo{person}{Hongyu Zhang}, \bibinfo{person}{Dongmei Zhang}, {and} \bibinfo{person}{Wenqiang Zhang}.} \bibinfo{year}{2021}\natexlab{}.
\newblock \showarticletitle{On the Evaluation of Commit Message Generation Models: An Experimental Study}. In \bibinfo{booktitle}{\emph{{IEEE} International Conference on Software Maintenance and Evolution, {ICSME} 2021, Luxembourg, September 27 - October 1, 2021}}. \bibinfo{publisher}{{IEEE}}, \bibinfo{pages}{126--136}.
\newblock


\bibitem[Tao et~al\mbox{.}(2024)]%
        {Tao_tosem2024}
\bibfield{author}{\bibinfo{person}{Wei Tao}, \bibinfo{person}{Yucheng Zhou}, \bibinfo{person}{Yanlin Wang}, \bibinfo{person}{Hongyu Zhang}, \bibinfo{person}{Haofen Wang}, {and} \bibinfo{person}{Wenqiang Zhang}.} \bibinfo{year}{2024}\natexlab{}.
\newblock \showarticletitle{{KADEL:} Knowledge-Aware Denoising Learning for Commit Message Generation}.
\newblock \bibinfo{journal}{\emph{{ACM} Trans. Softw. Eng. Methodol.}} \bibinfo{volume}{33}, \bibinfo{number}{5} (\bibinfo{year}{2024}), \bibinfo{pages}{133:1--133:32}.
\newblock


\bibitem[Tian et~al\mbox{.}(2022)]%
        {tian_icse2022}
\bibfield{author}{\bibinfo{person}{Yingchen Tian}, \bibinfo{person}{Yuxia Zhang}, \bibinfo{person}{Klaas{-}Jan Stol}, \bibinfo{person}{Lin Jiang}, {and} \bibinfo{person}{Hui Liu}.} \bibinfo{year}{2022}\natexlab{}.
\newblock \showarticletitle{What Makes a Good Commit Message?}. In \bibinfo{booktitle}{\emph{44th {IEEE/ACM} 44th International Conference on Software Engineering, {ICSE} 2022, Pittsburgh, PA, USA, May 25-27, 2022}}. \bibinfo{publisher}{{ACM}}, \bibinfo{pages}{2389--2401}.
\newblock


\bibitem[Touvron et~al\mbox{.}(2023)]%
        {touvron2023llama}
\bibfield{author}{\bibinfo{person}{Hugo Touvron}, \bibinfo{person}{Louis Martin}, \bibinfo{person}{Kevin Stone}, \bibinfo{person}{Peter Albert}, \bibinfo{person}{Amjad Almahairi}, \bibinfo{person}{Yasmine Babaei}, \bibinfo{person}{Nikolay Bashlykov}, \bibinfo{person}{Soumya Batra}, \bibinfo{person}{Prajjwal Bhargava}, \bibinfo{person}{Shruti Bhosale}, {et~al\mbox{.}}} \bibinfo{year}{2023}\natexlab{}.
\newblock \showarticletitle{Llama 2: Open foundation and fine-tuned chat models}.
\newblock \bibinfo{journal}{\emph{arXiv preprint arXiv:2307.09288}} (\bibinfo{year}{2023}).
\newblock


\bibitem[van Hal et~al\mbox{.}(2019)]%
        {van2019generating}
\bibfield{author}{\bibinfo{person}{SRP van Hal}, \bibinfo{person}{Mathieu Post}, {and} \bibinfo{person}{Kasper Wendel}.} \bibinfo{year}{2019}\natexlab{}.
\newblock \showarticletitle{Generating commit messages from git diffs}.
\newblock \bibinfo{journal}{\emph{arXiv preprint arXiv:1911.11690}} (\bibinfo{year}{2019}).
\newblock


\bibitem[V{\'{a}}squez et~al\mbox{.}(2015)]%
        {Vasquez_icse2015}
\bibfield{author}{\bibinfo{person}{Mario~Linares V{\'{a}}squez}, \bibinfo{person}{Luis~Fernando Cortes{-}Coy}, \bibinfo{person}{Jairo Aponte}, {and} \bibinfo{person}{Denys Poshyvanyk}.} \bibinfo{year}{2015}\natexlab{}.
\newblock \showarticletitle{ChangeScribe: {A} Tool for Automatically Generating Commit Messages}. In \bibinfo{booktitle}{\emph{37th {IEEE/ACM} International Conference on Software Engineering, {ICSE} 2015, Florence, Italy, May 16-24, 2015, Volume 2}}. \bibinfo{publisher}{{IEEE} Computer Society}, \bibinfo{pages}{709--712}.
\newblock


\bibitem[Wang et~al\mbox{.}(2021b)]%
        {DBLP:journals/tosem/WangXLHWG21}
\bibfield{author}{\bibinfo{person}{Haoye Wang}, \bibinfo{person}{Xin Xia}, \bibinfo{person}{David Lo}, \bibinfo{person}{Qiang He}, \bibinfo{person}{Xinyu Wang}, {and} \bibinfo{person}{John Grundy}.} \bibinfo{year}{2021}\natexlab{b}.
\newblock \showarticletitle{Context-aware Retrieval-based Deep Commit Message Generation}.
\newblock \bibinfo{journal}{\emph{{ACM} Trans. Softw. Eng. Methodol.}} \bibinfo{volume}{30}, \bibinfo{number}{4} (\bibinfo{year}{2021}), \bibinfo{pages}{56:1--56:30}.
\newblock


\bibitem[Wang et~al\mbox{.}(2021c)]%
        {Wang_tosem2021}
\bibfield{author}{\bibinfo{person}{Haoye Wang}, \bibinfo{person}{Xin Xia}, \bibinfo{person}{David Lo}, \bibinfo{person}{Qiang He}, \bibinfo{person}{Xinyu Wang}, {and} \bibinfo{person}{John Grundy}.} \bibinfo{year}{2021}\natexlab{c}.
\newblock \showarticletitle{Context-aware Retrieval-based Deep Commit Message Generation}.
\newblock \bibinfo{journal}{\emph{{ACM} Trans. Softw. Eng. Methodol.}} \bibinfo{volume}{30}, \bibinfo{number}{4} (\bibinfo{year}{2021}), \bibinfo{pages}{56:1--56:30}.
\newblock


\bibitem[Wang et~al\mbox{.}(2023)]%
        {Wang_ase2023}
\bibfield{author}{\bibinfo{person}{Liran Wang}, \bibinfo{person}{Xunzhu Tang}, \bibinfo{person}{Yichen He}, \bibinfo{person}{Changyu Ren}, \bibinfo{person}{Shuhua Shi}, \bibinfo{person}{Chaoran Yan}, {and} \bibinfo{person}{Zhoujun Li}.} \bibinfo{year}{2023}\natexlab{}.
\newblock \showarticletitle{Delving into Commit-Issue Correlation to Enhance Commit Message Generation Models}. In \bibinfo{booktitle}{\emph{38th {IEEE/ACM} International Conference on Automated Software Engineering, {ASE} 2023, Luxembourg, September 11-15, 2023}}. \bibinfo{publisher}{{IEEE}}, \bibinfo{pages}{710--722}.
\newblock


\bibitem[Wang et~al\mbox{.}(2021a)]%
        {Wang_emnlp2021}
\bibfield{author}{\bibinfo{person}{Yue Wang}, \bibinfo{person}{Weishi Wang}, \bibinfo{person}{Shafiq~R. Joty}, {and} \bibinfo{person}{Steven C.~H. Hoi}.} \bibinfo{year}{2021}\natexlab{a}.
\newblock \showarticletitle{CodeT5: Identifier-aware Unified Pre-trained Encoder-Decoder Models for Code Understanding and Generation}. In \bibinfo{booktitle}{\emph{Proceedings of the 2021 Conference on Empirical Methods in Natural Language Processing, {EMNLP} 2021, Virtual Event / Punta Cana, Dominican Republic, 7-11 November, 2021}}. \bibinfo{publisher}{Association for Computational Linguistics}, \bibinfo{pages}{8696--8708}.
\newblock


\bibitem[Weyssow et~al\mbox{.}(2025)]%
        {Weyssow_tosem2025}
\bibfield{author}{\bibinfo{person}{Martin Weyssow}, \bibinfo{person}{Xin Zhou}, \bibinfo{person}{Kisub Kim}, \bibinfo{person}{David Lo}, {and} \bibinfo{person}{Houari Sahraoui}.} \bibinfo{year}{2025}\natexlab{}.
\newblock \showarticletitle{Exploring Parameter-Efficient Fine-Tuning Techniques for Code Generation with Large Language Models}.
\newblock \bibinfo{journal}{\emph{ACM Trans. Softw. Eng. Methodol.}} (\bibinfo{date}{Jan.} \bibinfo{year}{2025}).
\newblock
\showISSN{1049-331X}


\bibitem[Whitehead(2007)]%
        {Whitehead2007CollaborationIS}
\bibfield{author}{\bibinfo{person}{Jim Whitehead}.} \bibinfo{year}{2007}\natexlab{}.
\newblock \showarticletitle{Collaboration in Software Engineering: A Roadmap}.
\newblock \bibinfo{journal}{\emph{Future of Software Engineering (FOSE '07)}} (\bibinfo{year}{2007}), \bibinfo{pages}{214--225}.
\newblock


\bibitem[Wilcoxon(1992)]%
        {wilcoxon1992individual}
\bibfield{author}{\bibinfo{person}{Frank Wilcoxon}.} \bibinfo{year}{1992}\natexlab{}.
\newblock \showarticletitle{Individual comparisons by ranking methods}.
\newblock In \bibinfo{booktitle}{\emph{Breakthroughs in statistics: Methodology and distribution}}. \bibinfo{publisher}{Springer}, \bibinfo{pages}{196--202}.
\newblock


\bibitem[Wu et~al\mbox{.}(2024)]%
        {wu2024commit}
\bibfield{author}{\bibinfo{person}{Yifan Wu}, \bibinfo{person}{Ying Li}, {and} \bibinfo{person}{Siyu Yu}.} \bibinfo{year}{2024}\natexlab{}.
\newblock \showarticletitle{Commit Message Generation via ChatGPT: How Far Are We?}. In \bibinfo{booktitle}{\emph{Proceedings of the 2024 IEEE/ACM First International Conference on AI Foundation Models and Software Engineering}}. \bibinfo{pages}{124--129}.
\newblock


\bibitem[Xu et~al\mbox{.}(2019)]%
        {Xu_ijcai2019}
\bibfield{author}{\bibinfo{person}{Shengbin Xu}, \bibinfo{person}{Yuan Yao}, \bibinfo{person}{Feng Xu}, \bibinfo{person}{Tianxiao Gu}, \bibinfo{person}{Hanghang Tong}, {and} \bibinfo{person}{Jian Lu}.} \bibinfo{year}{2019}\natexlab{}.
\newblock \showarticletitle{Commit Message Generation for Source Code Changes}. In \bibinfo{booktitle}{\emph{Proceedings of the Twenty-Eighth International Joint Conference on Artificial Intelligence, {IJCAI} 2019, Macao, China, August 10-16, 2019}}. \bibinfo{publisher}{ijcai.org}, \bibinfo{pages}{3975--3981}.
\newblock


\bibitem[Xue et~al\mbox{.}(2024)]%
        {Xue_issta2024}
\bibfield{author}{\bibinfo{person}{Zhiyi Xue}, \bibinfo{person}{Liangguo Li}, \bibinfo{person}{Senyue Tian}, \bibinfo{person}{Xiaohong Chen}, \bibinfo{person}{Pingping Li}, \bibinfo{person}{Liangyu Chen}, \bibinfo{person}{Tingting Jiang}, {and} \bibinfo{person}{Min Zhang}.} \bibinfo{year}{2024}\natexlab{}.
\newblock \showarticletitle{LLM4Fin: Fully Automating LLM-Powered Test Case Generation for FinTech Software Acceptance Testing}. In \bibinfo{booktitle}{\emph{Proceedings of the 33rd {ACM} {SIGSOFT} International Symposium on Software Testing and Analysis, {ISSTA} 2024, Vienna, Austria, September 16-20, 2024}}. \bibinfo{publisher}{{ACM}}, \bibinfo{pages}{1643--1655}.
\newblock


\bibitem[Yang et~al\mbox{.}(2024)]%
        {Yang_tse2024}
\bibfield{author}{\bibinfo{person}{Guang Yang}, \bibinfo{person}{Yu Zhou}, \bibinfo{person}{Xiang Chen}, \bibinfo{person}{Xiangyu Zhang}, \bibinfo{person}{Terry~Yue Zhuo}, {and} \bibinfo{person}{Taolue Chen}.} \bibinfo{year}{2024}\natexlab{}.
\newblock \showarticletitle{Chain-of-Thought in Neural Code Generation: From and for Lightweight Language Models}.
\newblock \bibinfo{journal}{\emph{{IEEE} Trans. Software Eng.}} \bibinfo{volume}{50}, \bibinfo{number}{9} (\bibinfo{year}{2024}), \bibinfo{pages}{2437--2457}.
\newblock


\bibitem[Yao et~al\mbox{.}(2023)]%
        {Yao_iclr2023}
\bibfield{author}{\bibinfo{person}{Shunyu Yao}, \bibinfo{person}{Jeffrey Zhao}, \bibinfo{person}{Dian Yu}, \bibinfo{person}{Nan Du}, \bibinfo{person}{Izhak Shafran}, \bibinfo{person}{Karthik~R. Narasimhan}, {and} \bibinfo{person}{Yuan Cao}.} \bibinfo{year}{2023}\natexlab{}.
\newblock \showarticletitle{ReAct: Synergizing Reasoning and Acting in Language Models}. In \bibinfo{booktitle}{\emph{The Eleventh International Conference on Learning Representations, {ICLR} 2023, Kigali, Rwanda, May 1-5, 2023}}. \bibinfo{publisher}{OpenReview.net}.
\newblock


\bibitem[Zheng et~al\mbox{.}(2024)]%
        {DBLP:journals/corr/abs-2403-13372}
\bibfield{author}{\bibinfo{person}{Yaowei Zheng}, \bibinfo{person}{Richong Zhang}, \bibinfo{person}{Junhao Zhang}, \bibinfo{person}{Yanhan Ye}, \bibinfo{person}{Zheyan Luo}, {and} \bibinfo{person}{Yongqiang Ma}.} \bibinfo{year}{2024}\natexlab{}.
\newblock \showarticletitle{LlamaFactory: Unified Efficient Fine-Tuning of 100+ Language Models}.
\newblock \bibinfo{journal}{\emph{CoRR}}  \bibinfo{volume}{abs/2403.13372} (\bibinfo{year}{2024}).
\newblock
\showeprint[arXiv]{2403.13372}


\end{thebibliography}
\end{document}